\documentclass[useAMS,usenatbib,twocolumn]{mnras}
\usepackage{amsmath}
\usepackage{mathtools}
\usepackage{mathrsfs}
\usepackage{yhmath}
\usepackage{bigints}
\usepackage{graphicx}
\usepackage{dcolumn}
\usepackage{multicol}
\usepackage{bm}
\usepackage{epsfig}
\usepackage{amssymb,latexsym,mathrsfs}
\usepackage{graphicx}
\usepackage{enumitem}
\usepackage{color}
\usepackage{hyperref}
\usepackage{float}
\usepackage{natbib}
\usepackage[utf8]{inputenc}
\usepackage[T1]{fontenc}
\usepackage[dvipsnames]{xcolor}
\usepackage{url}
\usepackage[normalem]{ulem}
\usepackage{array}
\usepackage[caption=false]{subfig}
\usepackage{multirow}
\makeatletter
\def\hlinewd#1{%
\noalign{\ifnum0=`}\fi\hrule \@height #1 \futurelet
\reserved@a\@xhline}
\makeatother
\hypersetup{
    colorlinks=true,
    linkcolor=red,
    citecolor=blue,
} 
\def\apj{{ ApJ}}
\def\apjl{{ ApJL}}
\def\apjs{{ ApJS}}

\def\aap{{ A\&A}}

\def\mnras{{ MNRAS}}

\newcommand{\be}{\begin{equation}}
\newcommand{\ee}{\end{equation}}
\newcommand{\ba}{\begin{eqnarray}}
\newcommand{\ea}{\end{eqnarray}}

\title[Multi-messenger signatures of delayed choked jets in TDEs]
{Multi-messenger signatures of delayed choked jets in tidal disruption events} 

\author[Mukhopadhyay et al.]{
Mainak Mukhopadhyay$^{1}$\thanks{mkm7190@psu.edu}, Mukul Bhattacharya$^{1,2}$, Kohta Murase$^{1,3,4}$\\
${}^1$Department of Physics; Department of Astronomy \& Astrophysics; Center for Multimessenger Astrophysics,
Institute for Gravitation\\ and the Cosmos, The Pennsylvania State University, University Park, PA 16802, USA\\
${}^2$Department of Physics, Wisconsin IceCube Particle Astrophysics Center, University of Wisconsin, Madison, WI 53703, USA\\
${}^3$Center for Gravitational Physics and Quantum Information, Yukawa Institute for Theoretical Physics, Kyoto University, Kyoto,\\ Kyoto 606-8502, Japan\\
${}^4$School of Natural Sciences, Institute for Advanced Study, Princeton, NJ 08540, USA
} 

\begin{document}

\date{Accepted . Received ; in original form }

\pagerange{\pageref{firstpage}--\pageref{lastpage}} \pubyear{2024}

\maketitle

\label{firstpage}

\begin{abstract}
Recent radio observations and coincident neutrino detections suggest that some tidal disruption events (TDEs) exhibit late-time activities, relative to the optical emission peak, and these may be due to delayed outflows launched from the central supermassive black hole. We investigate the possibility that jets launched with a time delay of days to months, interact with a debris that may expand outwards. 
We discuss the effects of the time delay and expansion velocity on the outcomes of jet breakout and collimation. We find that a jet with an isotropic-equivalent luminosity of $\lesssim 5 \times 10^{45}\,{\rm erg/s}$ is likely to be choked for a delay time of $\sim 3$ months. 
We also study the observational signatures of such delayed choked jets. The jet-debris interaction preceding the breakout would lead to particle acceleration and the resulting synchrotron emission can be detected by current and near-future radio, optical and X-ray telescopes, and the expanding jet-driven debris could explain late-time radio emission. We discuss high-energy neutrino production in delayed choked jets, and the time delay can significantly alleviate the difficulty of the hidden jet scenario in explaining neutrino coincidences. 
\end{abstract}


\section{Introduction}
\label{sec:intro}
Tidal disruption events (TDEs) are well-known sites of high-energy astrophysical phenomena. A TDE occurs when a star approaches sufficiently close to a supermassive black hole (SMBH) and is subsequently torn apart by the tidal forces of the SMBH~(see e.g., \citealt{Rees:1988bf,Stone:2012uk,Komossa:2015qya}). TDEs are prime candidates for multi-messenger observations and have attracted dedicated studies in neutrinos and electromagnetic emission, especially in the infrared (IR), optical, ultraviolet (UV), X-ray and radio bands. In particular, the mass fallback and accretion rates associated with TDEs can be tracked using optical/UV and X-ray observations~\citep{Stern:2004xb,Gezari:2007bw,Cenko:2011ys,Chornock:2013jta}, while radio observations help in characterizing outflows that originate following TDEs~\citep{Alexander:2020xwb}.
%

%
The SMBH-accretion disk system can power relativistic jets, and therefore, can act as a central engine in TDEs~\citep{Giannios:2011it,DeColle:2012np}. The existence of multiple jetted TDEs such as Swift J1644+57~\citep{Bloom:2011xk,Burrows:2011dn}, Swift J2058+05~\citep{Cenko:2011ys}, Swift J1112-8238~\citep{Brown:2015amy} and AT2022cmc~\citep{Andreoni:2022afu}, has been inferred by very bright and variable gamma/X-ray emission. 
Their large isotropic-equivalent energies strongly suggest the presence of a relativistic beamed jet~\citep{Burrows:2011dn,Zauderer:2011mf}. At later times, radio to millimetre emission typically follows as a result of the jet interaction with the circumnuclear material. 
Jetted TDEs have been explored in detail in the literature (see a review by~\citealt{DeColle:2019wzp}, and references therein), yet there are unresolved questions pertaining to the jet launching process, emission mechanism and outflow composition.
%

%
Such jetted TDEs have been considered as possible sources of ultrahigh-energy cosmic rays (UHECRs)~\citep{Farrar:2008ex,Farrar:2014yla} and high-energy neutrinos~\citep{Murase:2008zzc,Wang:2011ip}. 
Recent detection of high-energy neutrino events with IceCube, coincident with three TDE candidates (AT2019dsg, AT2019fdr and AT2019aalc) has unravelled yet another multi-messenger channel to study them. AT2019dsg associated with IceCube-191001A~\citep{Stein:2020xhk} is from a TDE that originated from a quiescent SMBH. AT2019fdr associated with IceCube-200530A~\citep{Reusch:2021ztx} is hosted by an unobscured active galactic nucleus (AGN). The search for TDEs being accompanied by an infrared echo further led to the coincidence between AT2019aalc and IceCube-191119A~\citep{vanVelzen:2021zsm}. 
Different production sites of high-energy neutrinos have been discussed, including successful jets~\citep{Dai:2016gtz,Senno:2016bso,Lunardini:2016xwi,Liu:2020isi,Winter:2020ptf}, choked or hidden jets~\citep{Senno:2016bso,Zheng:2022kam}, hidden winds \citep{Murase:2020lnu,Winter:2022fpf}, accretion disks \citep{Hayasaki:2019kjy,Murase:2020lnu} and coronae \citep{Murase:2020lnu}. 
Associated cascade gamma rays are also calculated \citep{Murase:2020lnu,Yuan:2023cmd}.
TDEs are also regarded as a population of hidden neutrino sources that are dark in GeV-TeV gamma rays, which have been required by the recent neutrino and gamma-ray data~\citep{Murase:2015xka,Capanema:2020rjj}

%
The physics of jet propagation has been extensively explored in the literature of gamma-ray bursts~\citep[e.g.,][]{Bromberg:2011fg,Mizuta:2013yma,MB_2023}. 
Analogously, TDE jets may interact with a stellar debris envelope, which can be static~\citep{Loeb:1997dv}, expanding~\citep[e.g.,][]{Strubbe:2009qs} or even contracting~\citep{Metzger:2022lob}. The launching time of jets and their direction against the orbital plane of the debris is debated~\citep[e.g.,][]{Tchekovsky2014}, which leads to diversity in the jet-debris interaction.  
If optically thick winds or unbound debris serve as an envelope around the SMBH, the jets can be choked and resulting electromagnetic emission can be obscured~\citep{Wang:2015mmh}, belonging to a class of hidden neutrino sources~\citep{Murase:2015xka}. 
Such jets can be even delayed. Indeed, recent radio observations of TDEs have shown significant time delays in radio emission relative to the peak time of optical emission~\citep{Horesh:2021gvp,Cendes:2022dia}. Coincident neutrino events were also observed in IceCube at $150\,{\rm days}$, $393\,{\rm days}$ and $148\,{\rm days}$ post the optical peaks for AT2019dsg, AT2019fdr and AT2019aalc, respectively~\citep{Stein:2020xhk,Reusch:2021ztx,vanVelzen:2021zsm}. This implies the existence of late-time central engine activities, and such a delay can make the jets choked more easily and motivates studies on the impacts on jet propagation and observational consequences. 

In this work, we primarily investigate the dynamics of delayed TDE jets while they propagate through the expanding stellar debris. In particular, we focus on the feasibility of choked jets for the physical parameters,  
particularly the time delay associated with the launch of the jet ($t_{\rm lag}$) and the debris velocity ($v_{\rm deb}$). We also present resulting multi-messenger signatures, focusing on the synchrotron emission from both forward and reverse shock regions. In the latter case, we study the fast and slow cooling regimes to discuss the detectability for radio, optical and X-ray telescopes. We also discuss implications for high-energy neutrino observations using the internal shock model~\citep{Senno:2016bso}. 

This paper is organised as follows. In Section~\ref{sec:model}, we discuss the physical model that we adopt to study the evolution of the stellar debris, the jet and the cocoon. The criteria used to determine jet collimation and breakout are discussed in Section~\ref{sec:criterion}. We present our main results in Section~\ref{sec:results}, where we also discuss the effect of varying important parameters in our model. The multi-messenger emission from delayed choked jets in TDEs, especially in the electromagnetic and neutrino channels, are discussed in Section~\ref{sec:signatures}. We summarize our results and conclude in Section~\ref{sec:disc}. 

%
\section{Physical model}
\label{sec:model}
\begin{figure}
\includegraphics[width=0.48\textwidth]{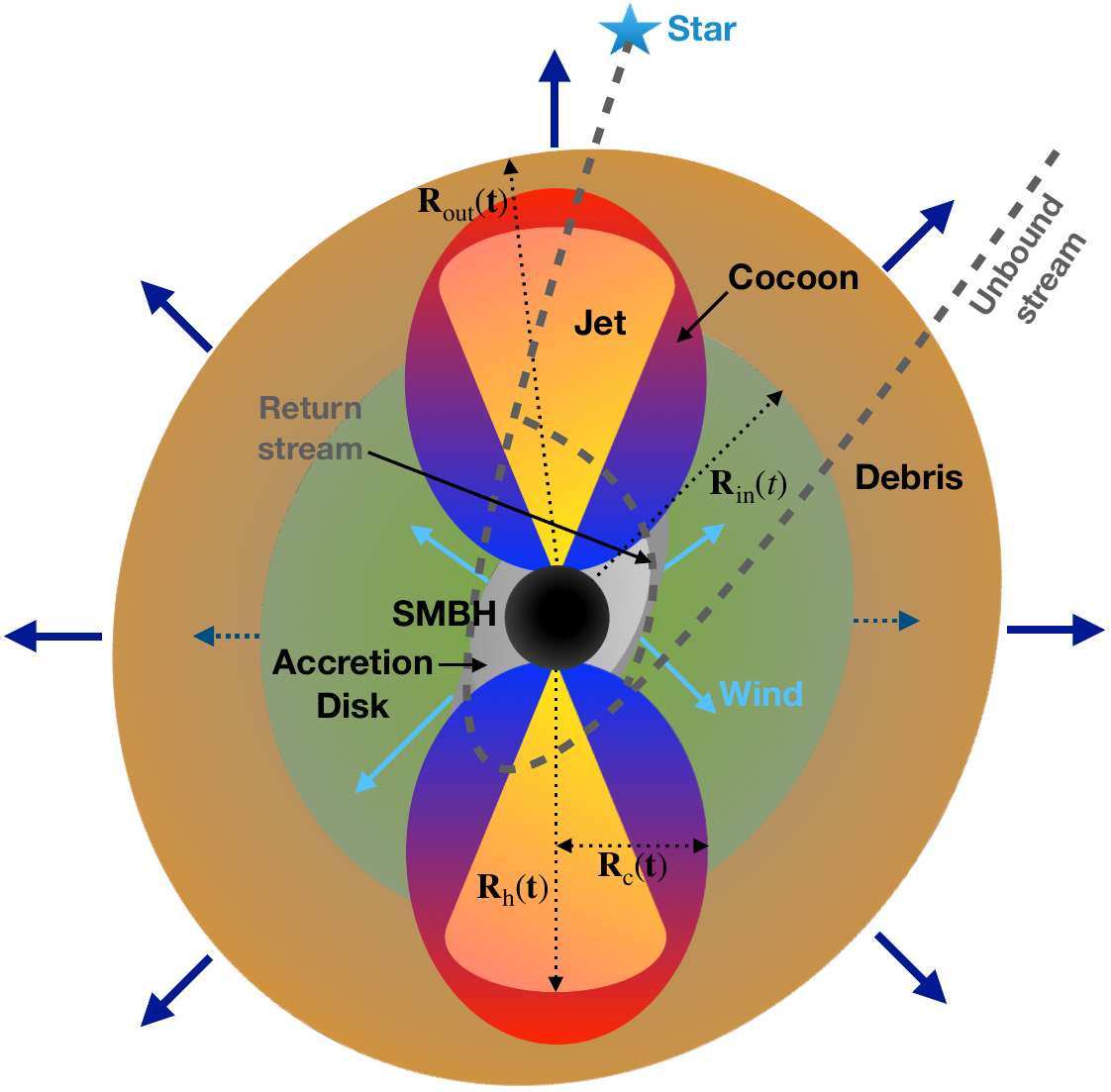}
\vspace{-0.3cm}
\caption{\label{fig:tde_schematic} Schematic diagram showing the evolution of the expanding aspherical debris (in brown), the relativistic jet and the mildly-relativistic cocoon, and low density wind-bubble region (in grayish-green color) post the tidal disruption of a star by the SMBH (in black) and the formation of an accretion disk (in gray). 
The inner/outer radius of the debris $R_{\rm in}(t)/R_{\rm out}(t)$, the jet-head radius $R_{\rm h}(t)$, and the cocoon radius $R_{\rm c}(t)$ are also labelled. The figure corresponds to the time snapshot at time $T$, where $t_{\rm coc}<T<t_{\rm br}$ or $t_{\rm coc}<T<t_{\rm fin}$, when the jet interacts with the debris (in brown) to form a cocoon (in blueish red). Note that $t_{\rm coc}$ is defined as the time when the cocoon is formed, $t_{\rm br}$ represents the time when the jet breaks out of the debris, and $t_{\rm fin}$ is the time until which the system is evolved. The trajectory of the unbound stream and the return stream are also shown with dashed dark grey lines.
}
\end{figure}
SMBHs are mostly found in the hearts of massive galaxies, including our own Milky Way. Tidal disruption drives the activity of some SMBHs, which can otherwise remain inactive for $\gtrsim 10^4$ years~\citep{vanVelzen:2018dwv}. The onset of TDE occurs once a star with mass $M_*$ and radius $R_*$ approaches the BH tidal radius, $R_{\rm tidal} = R_*(M_{\rm BH}/M_*)^{1/3}$. Throughout this work, we denote the total time since the TDE with $T = t + t_{\rm lag}$, where the time delay associated with jet launch is $t_{\rm lag}$ and the time since the jet launch is $t$, i.e. the jet is launched at $t = 0$.

In Figure~\ref{fig:tde_schematic}, we show a schematic for the model considered in this work. The infalling star is already disrupted by the SMBH (in black) and the disrupted stellar debris is assumed to form an accretion disk (in gray) and has an associated wind (shown with light blue arrows). This wind drives the aspherical expanding debris shown in brown (the dark blue arrows indicate expansion). The possible low-density wind-bubble region is also shown in grayish-green color. 
The schematic represents a time snapshot when the jet (shown in yellowish orange) has interacted with the debris which leads to the formation of the pressurised cocoon (shown in blueish red), that has an approximately ellipsoidal geometry.

Approximately, half of the disrupted stellar material falls back with timescales of $\sim10^6\, {\rm sec}$, to eventually form an accretion disk around the SMBH, whereas the other half becomes unbound~\citep{Evans:1989qe,Phinney1989}. 
For a SMBH with mass $M_{\rm BH} = 10^7 M_\odot$, the effective tidal disruption radius is given by $R_{\rm T} \approx f_T^{1/6} R_{\rm tidal}$ $\simeq (9.8 \times 10^{12}\,\text{cm})\ f_{\rm T,-1.1}^{1/6} M_{\rm BH,7}^{1/3}M_{*,0}^{2/3 - \xi}$. Here $M_{*}$ ($R_{*}$) is the mass (radius) of the tidally disrupted star, $f_{\rm T} \sim 0.02 - 0.3$ is a correction factor associated with the shape of the stellar density profile~\citep{Phinney1989,Piran:2015gha} and $\xi = 1 - {\rm ln}(R_*/R_\odot)/\rm ln(M_*/M_\odot)$ (for $M_* = M_\odot, \xi = 1$)\footnote{The factor $f_T$ appearing in the expression for the effective tidal-disruption radius $R_{\rm T}$ has some uncertainties associated with it~(see, e.g., \citealt{Golightly:2019jib,Ryu:2020huz}). We discuss this in more detail in Section~\ref{sec:disc}.}. 
Once a star gets disrupted close to $R_{\rm T}$, roughly half of its debris falls back with timescale, $t_{\rm fb} = 2\pi \sqrt{a_{\rm min}^3/(G M_{\rm BH})} \simeq (3.2 \times 10^6\,\text{s})\ f_{\rm T,-1.1}^{1/2} M_{\rm BH,7}^{1/2}M_{*,0}^{(1-3\xi)/2}$, where the semi-major axis of the orbit $a_{\rm min} \approx R_T^2/(2 R_*) \simeq (7.0 \times 10^{14}\,\text{cm})\ f_{\rm T,-1.1}^{1/3} M_{\rm BH,7}^{2/3}M_{*,0}^{1/3 - \xi}$. The debris eventually circularizes at a circularization radius $R_{\rm circ} \approx 2 R_{\rm T}$. A part of the debris eventually forms an accretion disk~\citep{Hayasaki:2012ia,Hayasaki:2015pxa,Bonnerot:2016krr,Bonnerot:2020pyz}.

Jetted TDEs constitute a well-motivated system in the literature owing to their associated luminous and variable gamma/X-ray emission. Since the focus of this work is to study the effects of delayed jet launch on the dynamics, we consider a simplified model for jet propagation. 
The stellar debris from the tidally disrupted star forms a thick circumnuclear envelope around the SMBH. For simplicity, we assume isotropic distribution for this envelope (at least along the direction of jet propagation), which we hereby refer to as the debris. It is important to note that this is a simplified assumption as we model the surrounding debris in an effectively isotropic time-averaged manner. 
This is not the case in general, since following stream-stream collisions at $T \sim t_{\rm fb}$, the debris becomes anisotropic and may have a clumpy structure. The orientation of BH spin and the disk need not be aligned, so the jet effectively interacts with the debris even if the debris has a torus-like geometry.

Here we are interested in studying the effect of time delay associated with jet launching, which typically exceeds $t_{\rm fb}$. For simplicity, we assume that the debris expands with a constant velocity and the bubble density is negligible compared to the density of the expanding debris. Thus, for all purposes considered here, the effect of the wind bubble on the jet propagation can be ignored. The density of the expanding debris is assumed to be,
\be
\label{eq:mdotw}
\rho (r) = \mathcal{N} \frac{M_{\rm deb}}{4 \pi R_{\rm out}^3} 
\left\{
\begin{array}{ll}
\bigg( r/R_{\rm out} \bigg)^{-2}, & r \geq R_{\rm fb} \vspace{0.3cm} \\
\bigg( R_{\rm fb}/R_{\rm out} \bigg)^{-2} \bigg( r/R_{\rm fb} \bigg)^{-\delta}, & r < R_{\rm fb} \\
\end{array}
\right., 
\ee
where $M_{\rm deb} = \eta_{\rm fb} M_*$, $R_{\rm out} = v_{\rm deb} T$, and $\delta \sim 0 - 1$. The normalization $\mathcal{N}$ is chosen such that, $\int_{R_{\rm in}(t)}^{R_{\rm out}(t)} \rho(r)\ dr = M_{\rm deb}$. The fallback radius $R_{\rm fb}$ evolves as
\be
R_{\rm fb}  = 
\left\{
\begin{array}{ll}
R_{\rm in}(T = 0), & T < t_{\rm fb} \vspace{0.3cm} \\
R_{\rm in}(T = 0) + v_{\rm deb} (T - t_{\rm fb}), & T \geq t_{\rm fb} \\
\end{array}
\right., 
\ee
where $R_{\rm in}(T = 0) = R_{\rm circ}$. Note that the time since tidal disruption is $T = t + t_{\rm lag}$, where $t$ is the time since jet launch. The density profile is valid for $T >> t_{\rm fb}$ which implies $t_{\rm lag} \sim 10^{6} - 10^{8}\,{\rm s}$. We assume $\delta = 1$ for this work.

The density profile of the debris extends from an inner radius $R_{\rm in}$(T) to an outer radius $R_{\rm out}$(T). The initial inner radius is set to the circularization radius, $R_{\rm in} (T=0) = R_{\rm circ}$, and is fixed until the fallback time. For $t>t_{\rm fb}$, the inner radius moves outwards as $R_{\rm in} (T) = R_{\rm circ} (T/t_{\rm fb})$. The outer radius of the debris is also assumed to start at $R_{\rm circ}$ and its evolution at later times is given by $R_{\rm out} (T) = v_{\rm deb} T$. In our simplified model, we adopt an expansion velocity of the debris within the range $v_{\rm deb} \sim 0.01-0.1\, c$, due to its inherent uncertainty. We consider $v_{\rm deb} \lesssim 0.1c$ because too energetic expanding outflows violate the radio data (although it is subject to uncertainty in the energy fraction carried by electrons)~\citep{Matsumoto:2021qqo}. We also expect that the velocity is not far from the escape velocity, $v_{\rm esc} (r = a_{\rm min}) = (2 GM/a_{\rm min})^{1/2} \sim 0.06c$. 

\subsection{Jet propagation in expanding ejecta}
\label{subsec:jetpropej}
The jet gets launched by the SMBH at $t = 0$ and has an associated time lag ($t_{\rm lag}$) relative to the time when the tidal disruption occurs. We assume that the jet is launched from the Schwarzschild radius $R_{\rm s} = 2GM_{\rm BH}/c^2$. After the jet is launched, the evolution of the jet-head is determined by, $\dot{R}_{\rm h} = c \beta_{\rm h}$ where, $R_{\rm h}$ is the vertical distance of the jet-head from the central engine and $\beta_{\rm h}c$ is the jet-head velocity. We use the subscript `h' to denote the jet-head quantities. The jet-head velocity is determined by the ram pressure balance between the shocked jet and the shocked envelope~\citep{Matzner:1998mg,Bromberg:2011fg},
\be
\label{beta_h}
\beta_{\rm h} = \frac{\beta_{\rm j} - \beta_{\rm a}}{1+\tilde{L}_{\rm c}^{-1/2}}+\beta_{\rm a}\,,
\ee
where, $\beta_{\rm j}c$ is the bulk velocity of the jet, $\beta_{\rm a}c$ is the velocity of the ambient medium and is set to $\beta_{\rm a}c = v_{\rm deb}$.
\begin{figure}
\includegraphics[width=0.49\textwidth]{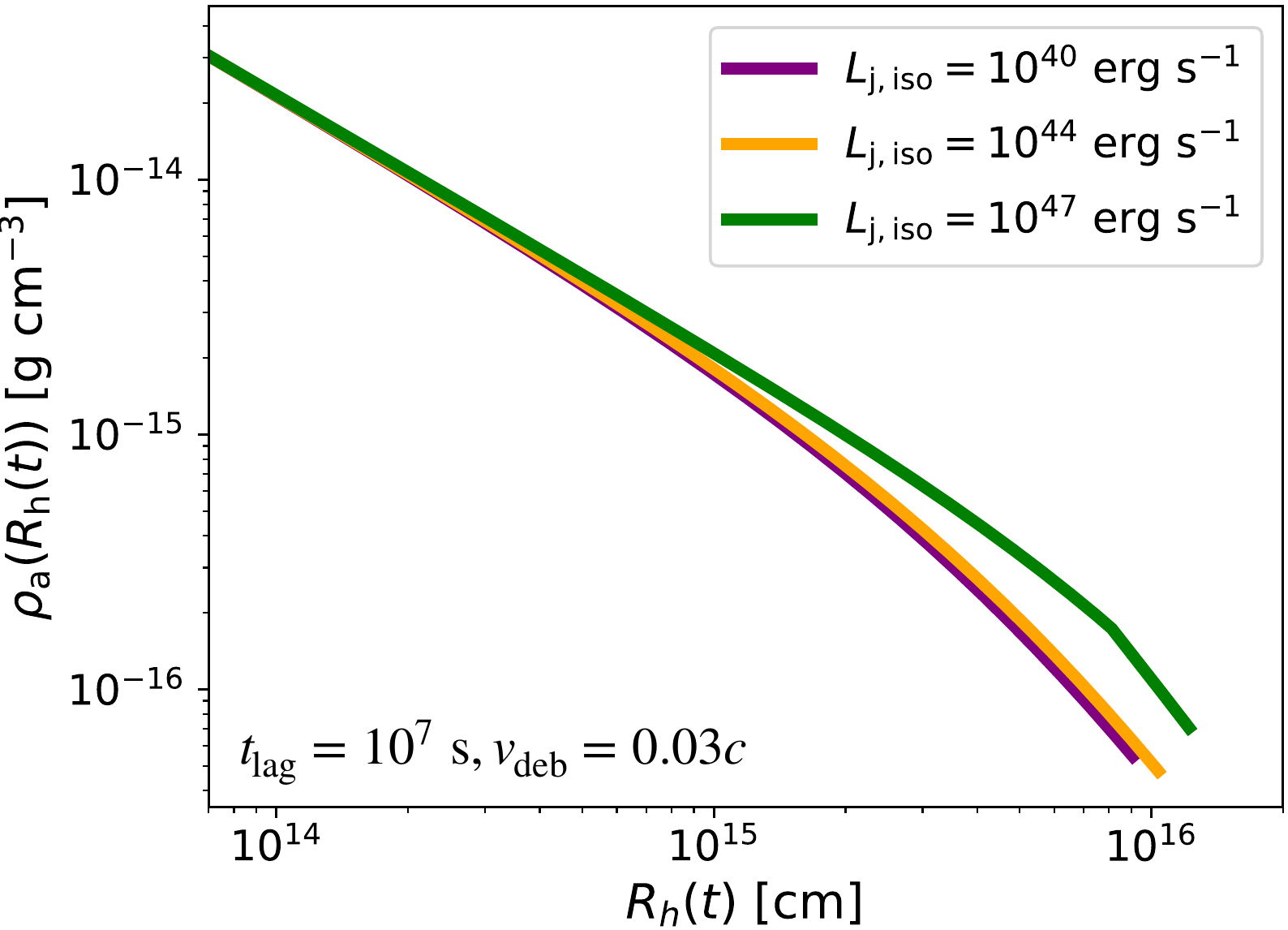}
\vspace{-0.3cm}
\caption{\label{fig:rho} Time evolution of the ambient medium density $\rho_{\rm a}(t)$ profile for different values of $L_{\rm j,iso}$. The evolution is terminated once the jet breaks out. The time delay is assumed to be $t_{\rm lag} = 10^7$s and the velocity of the expanding debris is taken to be $v_{\rm deb} = 0.03c$.
}
\end{figure}
The ratio of the energy density in the jet and the ambient medium can be defined as
\be
\label{eq:ltilde}
\tilde{L} = \frac{L_{\rm j}}{\Sigma_{\rm j}(t) \rho_{\rm a} (t) c^3 \Gamma_{\rm a}^2}\,.
\ee
In the above equation, $L_{\rm j}$ is the luminosity of the bipolar jet and the jet-head cross-section is $\Sigma_{\rm j}(t) = \pi R_h^2(t) \theta_{\rm j}^2(t)$, where $\theta_{\rm j}(t)$ is the opening angle of the jet. The jet opening angle plays an important role in collimation and breakout of the jet. The Lorentz factor of the ambient medium is given by $\Gamma_{\rm a} = 1/\sqrt{1-\beta_{\rm a}^2}$. The isotropic-equivalent luminosity of the jet is obtained from $L_{\rm j,iso} = L_{\rm j}/(0.5\ \theta_{\rm j}^2)$. The corresponding radiation component is given by $L_{\gamma, \rm iso} = \epsilon_\gamma L_{\rm j,iso}$, where $\epsilon_\gamma \sim 0.1$ is the efficiency for conversion of the jet luminosity to radiation. Finally, the jet pressure is given by $P_{\rm j}(t) = L_{\rm j}/(\Sigma_{\rm j}(t) c)$.

Following~\citet{Hamidani:2020krf}, we consider a calibrated value $\tilde{L}_{\rm c} = N_{\rm s}^2 \tilde{L}$, where the calibration factor $N_{\rm s} = N_{\rm s,0}(1-\beta_{\rm a})/\big( (1+N_{\rm s,0}\tilde{L}^{1/2})(1-\beta_{\rm a}^2)^{1/2}  \big)$ is used to match the jet breakout time obtained from numerical simulations with that given by known analytical estimates. 
However, currently we do not have sufficient number of TDE simulations that can be used to calibrate $\tilde{L}$. Given that the physical mechanism for jet launch and propagation in TDEs is similar to the jets from collapsars or BNS mergers, here we assume $N_s \approx 0.35$ for the fiducial case~\citep{Hamidani:2020krf}. 

The density of the ambient medium is obtained from $\rho_{\rm a} = \rho_{\rm deb}(r = R_h)$. Figure~\ref{fig:rho} shows the evolution of the ambient medium density profiles for different values of $L_{\rm j,iso}\sim10^{40-47}\,{\rm erg/s}$\footnote{Note that the broad range for the jet luminosity chosen here is to include the limiting possibilities. Besides, the jet efficiency can also be variable~\citep{Narayan:2021qfw}.}
We note that the ambient medium density does not vary significantly with an increase in the isotropic-equivalent luminosity from $10^{40}\,{\rm erg/s}$ (purple curve) to $10^{47}\,{\rm erg/s}$ (green curve) and decreases monotonically. For $L_{\rm j, iso} = 10^{47}\,{\rm erg/s}$, we clearly see the change in the power-law behaviour ($\rho_a \propto R_{\rm h}^{-2}$) once the jet-head radius exceeds the fall-back radius, that is $R_{\rm h}(t) > R_{\rm fb}(t)$, around $R_{\rm h} \sim 9 \times 10^{15}\,{\rm cm}$.

\subsection{Origin of delay ($t_{\rm lag}$) in jet launching}
\label{subsec:delay}
Recent observations of TDEs have shown a delay in radio emission relative to the time of optical discovery. For instance, ASASSN-15oi exhibited a radio emission peak $\sim180\, {\rm days}$ post its optical discovery~\citep{Horesh:2021gvp}. Similarly, radio emission from AT2018hyz was observed $\sim 970\, {\rm days}$ after its optical detection~\citep{Cendes:2022dia}. The origin of such delayed radio emission can be due to several possible effects:
\begin{enumerate}
\item A relativistic off-axis jet gets launched at the time of tidal disruption (see e.g., ~\citealt{Giannios:2011it,Mimica:2015qka,Generozov:2016oon}). However, as shown in~\cite{Cendes:2022dia}, an off-axis relativistic jet cannot reasonably explain the initial time delay for radio emission and the subsequent rebrightening for TDEs such as ASASSN-15oi.
\item An outflow that initially propagates in a low density medium, then interacts with a medium that has a significant density enhancement~\citep[e.g.,][]{NakarGranot2007}.
\item \label{item:3} The outflow itself has a delay associated with its launch relative to the tidal disruption and optical/UV emission time. Stream-stream collisions in TDEs have been studied to understand the process of disk formation~\citep{Lu:2019hwv,Bonnerot:2019yjb,Bonnerot:2020pyz}. However, the relevant timescales and geometry associated with disk formation still remain uncertain. The formation of the accretion disk, which may depend on details of the circularization and envelope cooling can be delayed~\citep[e.g.,][]{Metzger:2022lob}. 

\item Even after the disk is formed, a state transition in the accretion disk, e.g., from the standard disk to the radiatively inefficient accretion flow, can happen at $\sim10^6-10^8$~s, depending on several factors such as the SMBH mass~\citep{Murase:2020lnu}. The transition time is estimated to be
\begin{equation}
t_{\rm RIAF}\sim 6\times{10}^6~{\rm s}~\alpha_{-1}^{-51/19}M_{\rm BH,7}^{35/19}{\mathcal H}^{-2}{(R_d/10R_S)}^{3/2},
\end{equation}
where $\alpha$ is the viscosity parameter, ${\mathcal H}$ is the normalized disk scale height and $R_d$ is the disk radius, which can be consistent with the time delay of neutrino events and radio detection. In addition, it is believed that the launching of relativistic jets requires a very strong magnetic field at the BH event horizon~\citep{Tchekovsky2014}. Although a dynamo effect in the accretion disk can generate sufficiently strong magnetic fields to power the jet~\citep{Liska:2018btr}, the accumulation of the magnetic flux onto the SMBH may be delayed~\citep{Tchekovsky2014,Kelly2014}.
\end{enumerate}

In this work, we mainly focus on the third scenario~\ref{item:3} where we define the delay associated with the jet launching at the SMBH as $t_{\rm lag}$. 
For our purposes here, we assume the delay in jet launching within the range $t_{\rm lag}\sim 10^6-10^8\, {\rm sec}$. It must be noted here that due to the simplified nature of our model, the dynamics of the model and the results presented are more accurate when $t_{\rm lag} \gtrsim t_{\rm fb}$.
%
\subsection{Jet-cocoon system}
\label{subsec:jetcocnsys}
As the jet-head encounters the debris, the ambient matter gets heated and moves sideways, which leads to the formation of a pressured cocoon around the jet. The cocoon pressure is dominated by radiation pressure and is given by
\be
\label{eq:pc}
P_{\rm c}(t) = \frac{E_{\rm c}}{3 V_{\rm c}}  = \frac{\eta}{4\pi R_{\rm c}(t)^2 R_{\rm h}(t)}  \int_{t_{\rm c}}^t d \tilde{t}\ L_{\rm j}(\tilde{t}) \Big(1-\beta_{\rm h}(\tilde{t})\Big)\,,
\ee
where, $R_{\rm c}(t)$ is the lateral radius of the cocoon, $t_{\rm c}$ is the time at which the jet-head reaches $R_{\rm in}$, and the cocoon formation starts. A steady energy inflow from the jet-head sustains this cocoon pressure. The parameter $\eta$ in equation~\eqref{eq:pc} is defined as the fraction of jet energy that is deposited into the cocoon while the jet propagates within the ejecta. Based on detailed numerical simulations, it is taken as $\sim 1/4$ for BNS mergers and $\sim 1/2$ for collapsars~(see e.g., \citealt{Hamidani:2019qyx}). As earlier, limited number of detailed TDE simulations prevents us from making a robust estimate and we hereby adopt $\eta = 1$. To compute the cocoon volume $V_c$ in equation~\eqref{eq:pc}, we assume an ellipsoidal geometry with a semi-major axis $(1/2) R_{\rm h}(t)$ and a semi-minor axis $R_{\rm c}(t)$. Note that this differs from a cylindrical geometry that was assumed in some previous works (e.g., \citealt{Bromberg:2011fg,Mizuta:2013yma}). 

For a numerical estimate, the integral in equation~\eqref{eq:pc} can be approximated with an average value method discussed in~\cite{Hamidani:2020krf}, where the integral is replaced by $L_{\rm j} (t)(1- \langle \beta_{\rm h} \rangle ) (T-t_{\rm lag})$. Here the time-averaged jet-head velocity $\langle \beta_{\rm h} \rangle$ is defined as
\be
\label{eq:avgint}
\langle \beta_h \rangle  = \frac{1}{c t} \big[R_{\rm h}(t) - R_{\rm h}(t=0) \big]\,.
\ee
The dynamics of the cocoon, in particular, the time evolution of its semi-minor axis is governed by $\dot{R}_{\rm c} = c \beta_{\rm c}$, where $\beta_c c$ is the lateral velocity of the cocoon and is given by~\citep{Hamidani:2019qyx},
\be
\beta_{\rm c} \approx \frac{1}{c}\sqrt{\frac{P_{\rm c}}{\rho_{\rm a}(t)}} + \frac{R_{\rm c}(t)}{R_{\rm out}(t)} \frac{v_{\rm w}}{c}\,.
\ee
The above expression for the cocoon's lateral velocity assumes vertical height of the cocoon to be the same as the jet-head position. This approximation becomes better as the system is evolved for longer. The initial cocoon radius is set as $R_c(t = t_{\rm coc}) = R_{\rm j}(t=t_{\rm coc}) = R_{\rm h}(t=t_{\rm coc}) \theta_0$, where $\theta_0=\theta_{\rm j}(t=0)$ is the initial jet opening angle, $R_{\rm j}(t)$ is the lateral radius of the jet-head and $t_{\rm coc}$ is the  time at which the jet-head reaches the inner radius of the debris, $R_{\rm h}(t=t_{\rm coc}) = R_{\rm in} (t_{\rm coc}+t_{\rm lag})$. The second term in the expression for $\beta_c$ is included to account for homologous expansion of the debris.

%
\section{Criterion for collimation and breakout of the jet}
\label{sec:criterion}
As the jet propagates through the debris, its interaction with the surrounding cocoon plays an important role in deciding the jet-head velocity as well as cross-section of the jet, and therefore the outcomes for jet collimation and breakout. 
In this section, we analyse whether the necessary conditions for collimation and breakout of the jet are achieved for the range of physical parameters that we consider in this study.
Numerous analytical~\citep{Blandford:1974zz,Begelman:1989jp,Meszaros:2001ms,Matzner:2002ti,Lazzati:2005xv,Bromberg:2011fg} and numerical investigations~\citep{Aloy:1999ai,Zhang:2003rp,Lazzati_2009,Mizuta:2008ch,Nagakura:2010zt,Mizuta:2013yma}, have been performed to study the propagation of hydrodynamic jets through both static and expanding external media. 
%
\subsection{Jet collimation}
\begin{figure}
\includegraphics[width=0.49\textwidth]{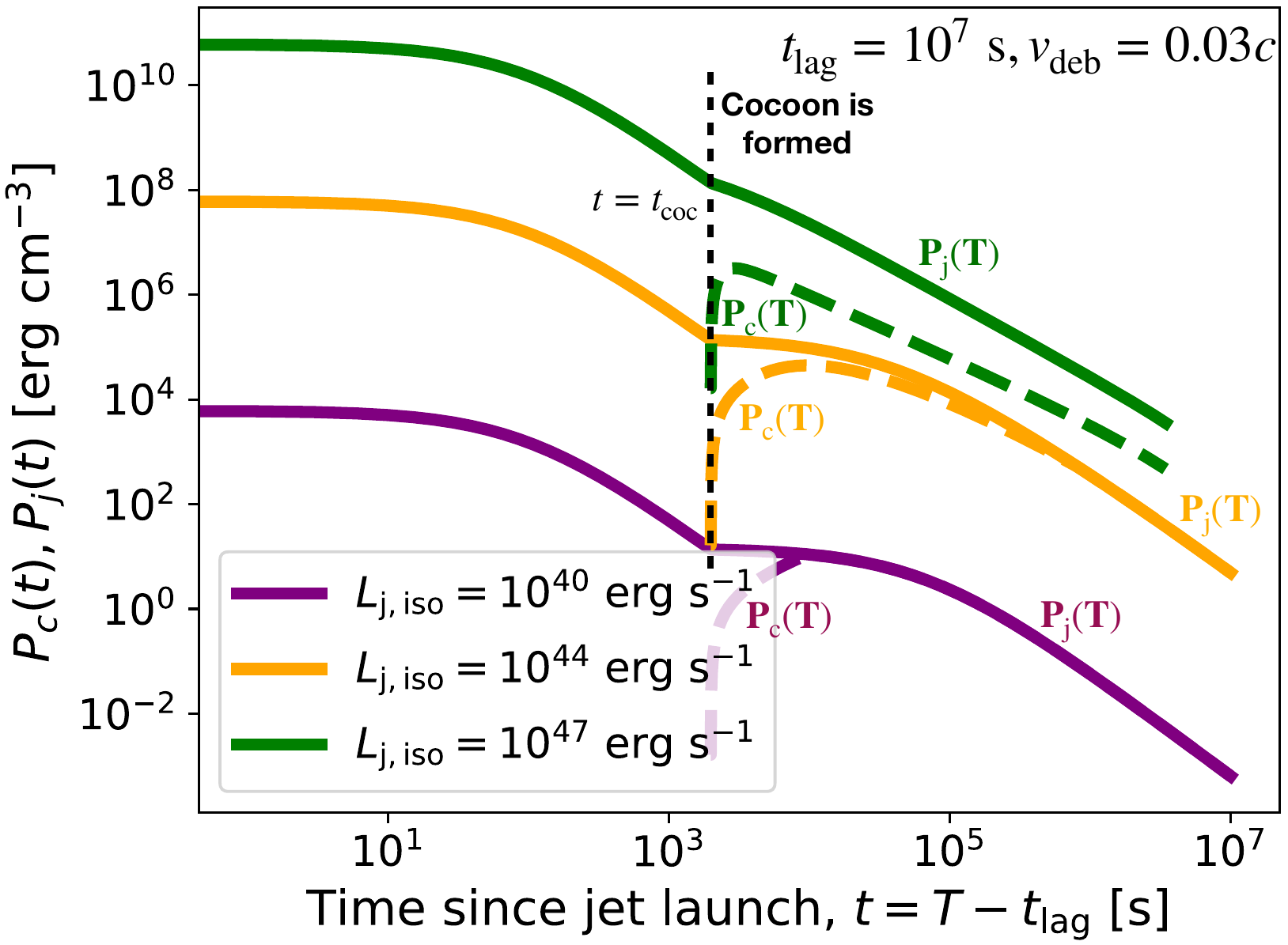}
\vspace{-0.3cm}
\caption{\label{fig:pcpjplots} 
Time evolution of $P_{\rm c}$ and $P_{\rm j}$ are shown for three different values of $L_{\rm j, iso}$ to illustrate the collimation and no collimation scenarios. The solid lines denote $P_{\rm j}(t)$, whereas $P_{\rm c}(t)$ is shown using the dashed lines. For $L_{\rm j, iso} = 10^{40}\,{\rm erg/s}$ (purple curves) and $L_{\rm j, iso} = 10^{44}\,{\rm erg/s}$ (orange curves), $P_{\rm c} = P_{\rm j}$ is eventually achieved leading to collimation of the jet. However, for $L_{\rm j, iso} = 10^{47}\,{\rm erg/s}$ (green curves), the jet breaks out before the ambient medium can collimate it. The instant of cocoon formation ($t=t_{\rm coc}$) is shown by the black dashed vertical line. In each case, the evolution is terminated once the jet breaks out of the stellar debris or when $t_{\rm fin}$ is reached.
}
\end{figure}
\citet{Bromberg:2011fg} showed that oblique shocks that form inside the relativistic outflow close to the jet base and converge on the jet-axis can collimate the outflow. The oblique shock formed at the jet base counterbalances the cocoon pressure $P_{\rm c}$. 
Upon collimation, the jet geometry changes from conical to cylindrical, and consequently, the jet-head cross-section decreases significantly. The cocoon height, and therefore its volume, increases leading to a lower cocoon pressure. Thus, for $\beta_{\rm h}$ larger than some critical value, $P_{\rm c}$ decreases to an extent such that it is no longer sufficient to collimate the jet. 
At a given time during its evolution, the collimation criteria for the jet is given by
\be
P_c(t) = P_j(t),\ \ {\rm for}\  R_{\rm in}(t) \leq R_h(t) \leq R_{\rm out}(t).
\ee
The radius at which the jet gets collimated by the cocoon is defined as, $R_{\rm coll} = R_h(t = t_{\rm coll})$, where $t_{\rm coll}$ is the time of collimation. 

In Figure~\ref{fig:pcpjplots}, we show the jet and cocoon pressure with solid and dashed lines, respectively, for three different values of $L_{\rm j, iso} = 10^{40},\, 10^{44},\, 10^{47}\,{\rm erg/s}$. In all the cases, the jet reaches the inner radius of the debris leading to the formation of the cocoon. This is marked by a dip in the jet pressure at $t =t_{\rm coc}$, corresponding to $t \sim 10^3\,{\rm s}$, and is shown by the dashed vertical line. The cocoon pressure begins to rise after this. Upon collimation, based on the criteria defined above, the jet and the cocoon pressure become equal and evolve in the same manner. The low-luminosity jets collimate early as is evident from the purple curves. As luminosity increases, the time required for the jet to be collimated also increases, as can be seen by comparing the purple and orange curves. For high-luminosity jets (green curves), collimation is not achieved for the duration of our simulations. This is expected since it is well-known that it is easier to collimate low-luminosity hydrodynamic jets. In Section~\ref{subsec:res_tlag}, we discuss the effect of $t_{\rm lag}$ on the collimation radius. We also note that the dip in $P_{\rm j}$ is less pronounced for $L_{\rm j,iso} = 10^{47}\,{\rm erg/s}$.

%
\subsection{Jet breakout}
\label{subsec:jetbr}
\begin{figure}
\includegraphics[width=0.47\textwidth]{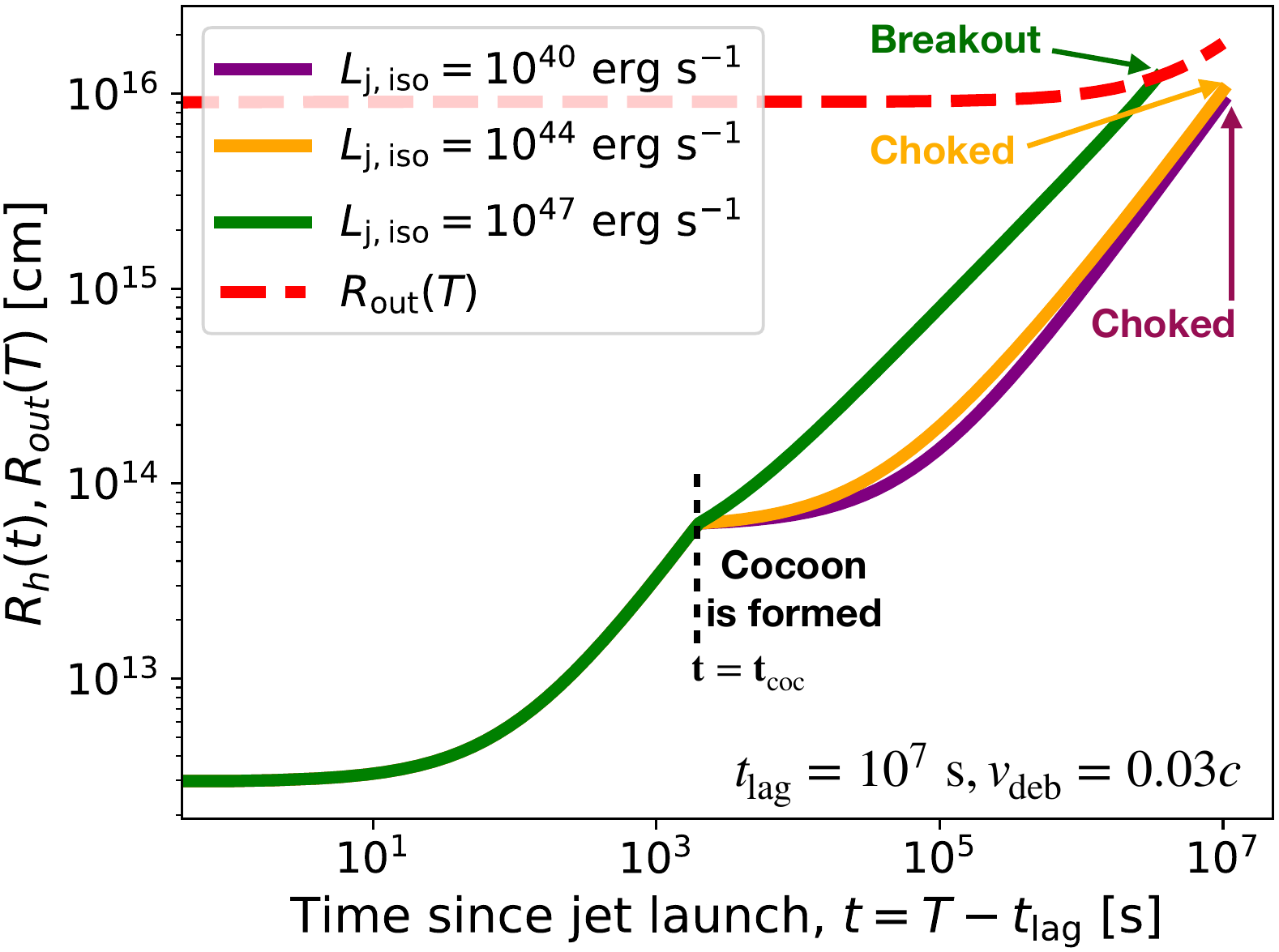}
\caption{\label{fig:rplots} Time evolution of the jet-head position $R_{\rm h} (t)$ (solid lines) and outer radius of the wind-driven debris $R_{\rm out}(t)$ (red dashed line). The evolution of jet-head position is shown for different values of $L_{\rm j,iso}$, where $L_{\rm j,iso} = 10^{40}\,{\rm erg/s}$ (purple curve) and $L_{\rm j,iso} = 10^{44}\,{\rm erg/s}$ (orange curve) jets are choked within the wind-driven debris, but the $L_{\rm j,iso} = 10^{47}\,{\rm erg/s}$ (green curve) jet breaks out. The time at which the jet-head reaches the inner radius of the wind-driven debris is shown by the black dashed line. The evolution is terminated once the jet breaks out or $t_{\rm fin}$ is reached.
}
\end{figure}
Similar to jet collimation, the jet breakout condition also depends on the interaction between the jet and the cocoon. Hydrodynamic jets may get choked inside the star if the jet isotropic-equivalent luminosity does not exceed a critical value~\citep{Meszaros:2001ms}. The jets which do not break out of the stellar envelope are known as choked jets. In case of a choked jet, the cocoon may still break out from the ejecta and lead to detectable electromagnetic emission. However, in this work we terminate our simulations at the time when the jet breaks out of the ejecta. 

Since the jet is powered by the SMBH, there exists a threshold energy that the latter needs to generate to push the jet out of the stellar envelope. The time for which the SMBH needs to be active to generate this threshold energy is $t_{\rm th} = t_{\rm br}-R_{\rm out}/c$. So, the minimal engine activity time $t_{\rm eng}$ needs to exceed $t_{\rm th}$ for the jet to successfully break out. The jet can get choked within the ejecta envelope due to two reasons:
(1) the isotropic jet power is less than the threshold luminosity required for breakout, or (2) the time for which the SMBH powers the jet is less than $t_{\rm th}$. 
We assume the jet to have broken out when
\be
\label{eq:breakoutcrit}
R_{\rm h} (t) \geq R_{\rm out}(t),\ {\rm for}\ t \gtrsim t_{\rm lag}.
\ee
The time at which the above criterion is satisfied is known as the \emph{jet breakout time} ($t_{\rm br}$). 

In Figure~\ref{fig:rplots}, we show the evolution of the jet-head radius $R_{\rm h} (t)$ for different values of $L_{\rm j,iso}\sim10^{40-47}\,{\rm erg/s}$. The evolution of $R_{\rm out} (t)$ for $v_{\rm deb} = 0.03c$ is also shown as a dashed red line in the figure. The jet is launched at $t = 0$ for all the cases shown. As discussed in Section~\ref{sec:model}, initially the jet propagates freely through the wind bubble which has negligible density. The jet-head radius for different luminosities evolve similarly until the jet-head reaches the inner radius of the debris. Following this, the cocoon forms and the jet-cocoon dynamics become important. The jet with higher luminosity ($10^{47}$ erg/s, green curve) breaks out much earlier than the other two cases of intermediate ($10^{44}$ erg/s, orange curve) and low ($10^{40}$ erg/s, purple curve) luminosities, which serve as examples of choked jets. The effects of $t_{\rm lag}$ on the time of breakout for the jets is discussed in Section~\ref{subsec:res_tlag}.

Here we provide an analytical estimate for the jet choking criterion from equation~\eqref{eq:breakoutcrit}, by solving  $R_{\rm h} (t) \leq R_{\rm out} (T)$. As the jet evolves until $t = t_{\rm dur}$, we determine this criterion at the final evolution time i.e. $T = t_{\rm dur} + t_{\rm lag}$. The outer radius of the debris, $R_{\rm out} (T) = v_{\rm deb} (t_{\rm dur}+t_{\rm lag})$. This can be expressed as,
\be
R_{\rm out} \simeq 1.8 \times 10^{16}\,{\rm cm}\, \bigg( \frac{\beta_{\rm deb}}{0.03}\bigg)\bigg( \frac{t_{\rm dur}}{10^{7}\rm s}\bigg) \bigg(\frac{\chi_{\rm lag}}{2}\bigg)\,,
\ee
where, $\chi_{\rm lag} = 1+t_{\rm lag}/t_{\rm dur}$ and $\beta_{\rm deb} = v_{\rm deb}/c$. The jet-head position can be estimated as (see \citealt{Bromberg:2011fg,Mizuta:2013yma,Harrison:2017jvs,Kimura:2018vvz} for details),
\be
\label{eq:rhchok}
\begin{split}
R_{\rm h} &\simeq 5.6 \times 10^{15}\,{\rm cm}\, \bigg( \frac{N_s}{0.35} \bigg)^{5/3} \bigg( \frac{L_{\rm j,iso}}{10^{44}\,{\rm erg/s}} \bigg)^{1/3} \bigg( \frac{M_{\rm deb}}{0.5\,M_{\odot}}\bigg)^{-1/3} \\
&\bigg( \frac{\theta_0}{0.17} \bigg)^{-2/3}\bigg( \frac{\beta_{\rm deb}}{0.03} \bigg)^{1/3} \bigg( \frac{t_{\rm dur}}{10^7\,{\rm s}}\bigg)^{4/3} \bigg( \frac{\chi_{\rm lag}}{2}\bigg)^{1/3}\,.
\end{split}
\ee
Here to obtain an analytical estimate on the critical luminosity for the jets to be choked, we do not consider the constant prefactor $\sim 32$ for collimated jets as done by \citet{Mizuta:2013yma}. This is because for higher values of $L_{\rm j,iso}$ choked jets are predominantly uncollimated. 
We also fix the normalization of the density profile using the fiducial case and do not include the time-dependent normalization term for simplicity. This is once again a valid estimate since $\mathcal{N} \sim 1$. Finally, the critical luminosity for choked jets is estimated to be,
\be
\label{eq:chokcrit}
\begin{split}
L_{\rm j,iso} &\lesssim 3.2\times10^{45}\,{\rm erg/s}\,\bigg( \frac{N_s}{0.35} \bigg)^{-5} \bigg( \frac{M_{\rm deb}}{0.5M_{\odot}}\bigg) \bigg( \frac{\theta_0}{0.17} \bigg)^{2} \\
&\bigg( \frac{\beta_{\rm deb}}{0.03} \bigg)^{2} \bigg( \frac{t_{\rm dur}}{10^7\,{\rm s}}\bigg)^{-1} \bigg( \frac{\chi_{\rm lag}}{2}\bigg)^{2}\,.
\end{split}
\ee
\section{Results}
\label{sec:results}
\begin{figure*}
\begin{center}
\subfloat[(a)] {\includegraphics[width=0.49\textwidth]{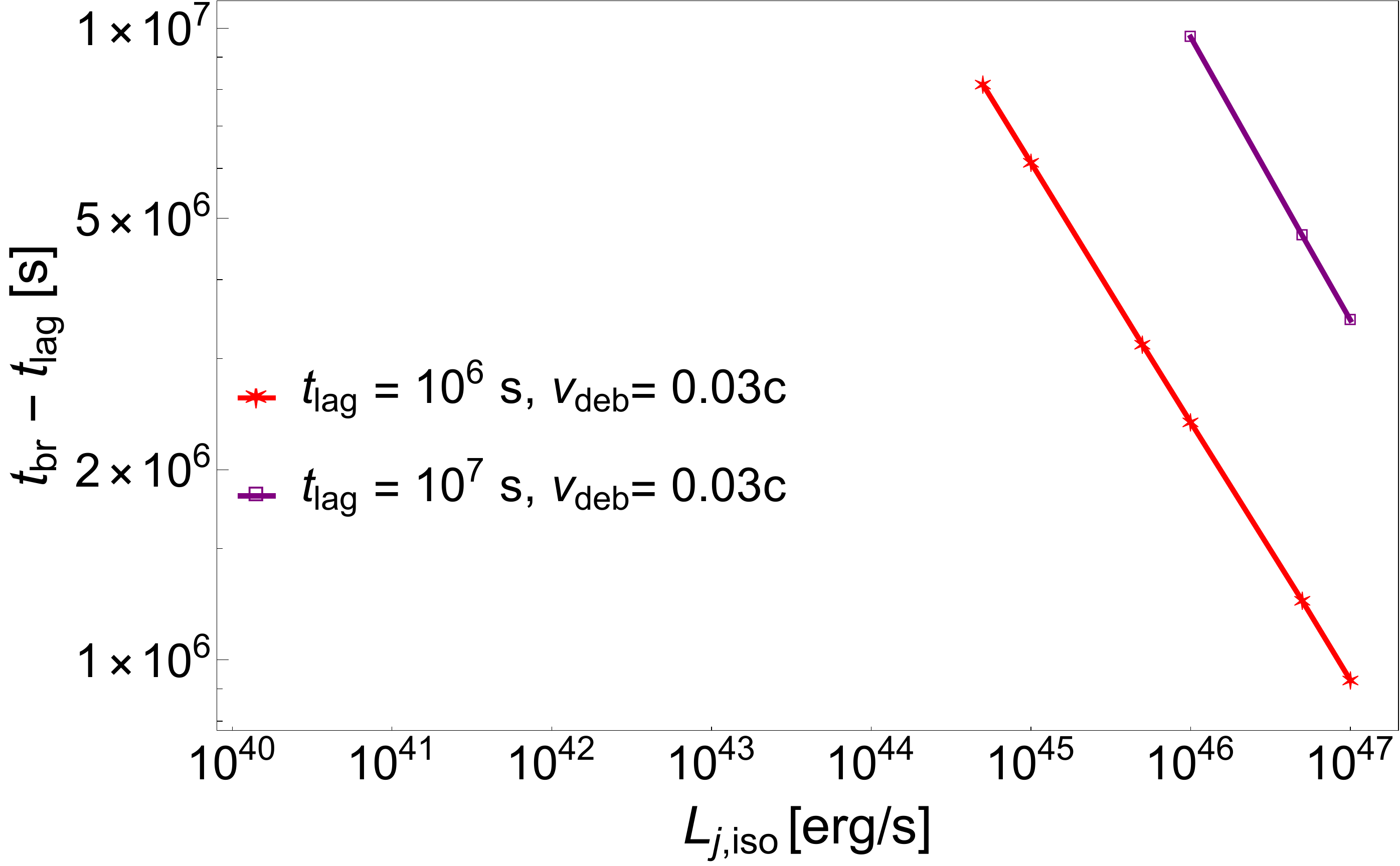}\label{fig:tbr_tlag}} \hfill
\subfloat[(b)]{\includegraphics[width=0.49\textwidth]{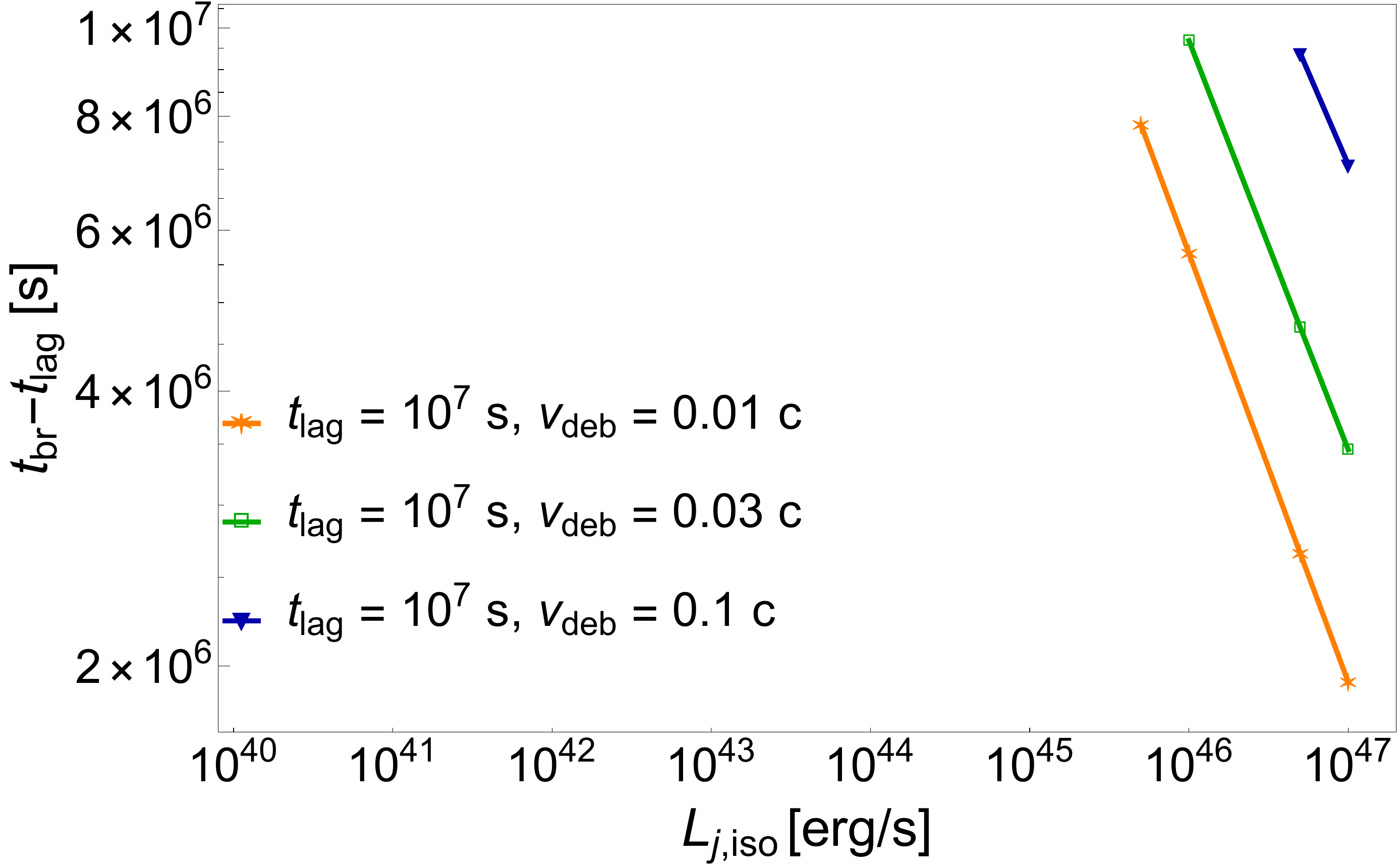}\label{fig:tbr_vw}} \hfill
\caption{\label{fig:res_tlag}  
The breakout time $t_{\rm br}$ is shown as a function of varying jet isotropic-equivalent luminosity $L_{\rm j,iso}$: (a) for two different values of $t_{\rm lag}$ with $v_{\rm deb} = 0.03c$ (for $t_{\rm lag} = 10^8\,{\rm s}$, none of the jets in our model breakout, as can be seen from Table~\ref{tab:res1} and hence that case is not shown in the figure), and (b) for three different values of $v_{\rm deb}$ with $t_{\rm lag} = 10^7\,{\rm s}$. The numerical data points corresponding to the values of $L_{\rm j,iso}$ that are not shown here imply that successful jet breakout does not occur for the entire duration of our system evolution.
}
\end{center}
\end{figure*}

In this section, we discuss the main results of this work. We are primarily interested in the collimation and choking conditions for the jet, given its interaction with the cocoon in the presence of the wind-driven debris. In this context, we study the occurrence of collimation and the jet breakout time ($t_{\rm br}$), for a given set of parameters of the physical system. Although the unavailability of detailed numerical simulations of jetted TDEs makes it difficult to fix parameters, we construct a well-motivated fiducial case and the main parameters varied are the time delay associated with jet launching $t_{\rm lag}$ and the expansion velocity of the debris $v_{\rm deb}$. These two parameters are not well constrained in TDE literature and the focus of this work is to study their effect on the dynamics of the system. Another interesting parameter to vary would be the initial jet opening angle $\theta_0$. For the fiducial case we fix it to $10^\circ$ to illustrate our results. Reducing $\theta_0$ slightly decreases the critical luminosity for choked jets, whereas increasing $\theta_0$ increases the same. For example, for $\theta_0 = 5^\circ$ we find $L_{\rm j,iso} \lesssim 10^{45}\,{\rm erg/s}$ jets get choked, while for $\theta_0 = 20^\circ$, $L_{\rm j,iso} \lesssim 5 \times 10^{46}\,{\rm erg/s}$ jets get choked corresponding to $v_{\rm deb} = 0.03c$ and $t_{\rm lag} = 10^7\,{\rm s}$. This can also approximated from equation~\eqref{eq:chokcrit}. Thus, we do not vary $\theta_0$ to obtain our results. We present our results for different values of $L_{\rm j,iso}$ ranging from $10^{40}\,{\rm erg/s}$ to $10^{47}\,{\rm erg/s}$. 

To obtain our results, we solve the evolution equations for the jet-head $R_{\rm h}(t)$, and the lateral radius of the cocoon $R_{\rm c}(t)$. The ejecta starts expanding at $T=0$, i.e., just after the TDE occurs. The jet, however, is assumed to have a delay in its launching from the central SMBH, as discussed in detail in Section~\ref{subsec:delay}. The jet-head emerges from the SMBH at $T=t_{\rm lag}$ (that is, $t = 0$), where $t_{\rm lag}$ is assumed to be within $\sim$few days to a few years, i.e., $t_{\rm lag}\sim 10^6-10^8\,{\rm s}$. It is easy to see that $t_{\rm lag} = 0$ would be the case where the jet is launched instantaneously and has no associated delay. Our simplified model does not apply in that scenario and provides reasonable estimates only when $t_{\rm lag} \gtrsim t_{\rm fb}$.

For the results shown in this work, we assume a fixed SMBH mass $M_{\rm BH} = 10^7 M_\odot$. We also fix the mass and radius of the infalling star to be $M_* = 1 M_\odot$ and $R_* = 1 R_\odot$, respectively.
We evolve the jet-cocoon system for a given parameter set until $T=t_{\rm fin}$. We assume that the jet is powered until the fiducial $t_{\rm lag}$ that is $\sim$ 3 months, which is observationally well-suited. Thus we set $t_{\rm dur} \sim 10^7\ \rm s$, which implies $t_{\rm fin} = 10^7\ \rm s + t_{\rm lag}$.
It is important to note that for the expanding debris and the associated delay in jet launching, it is unclear whether $t_{\rm dur} \sim 10^7\ \rm s$, but we use this approximation for the current work. Besides, recent observations corroborate that the electromagnetic signatures from TDEs, particularly the radio and neutrino emissions, are over the timescales of $\sim$ a few 100 days. The choice of the total evolution time is important since choosing a short timescale might result in getting spurious results whereby the jet is either not collimated or does not breakout of the ejecta, solely because the dynamical timescales of the physical system are longer. As a final remark, we note that it is also important to choose a small enough time-step to model the system evolution so as to not miss out on any of the important dynamics of the jet or the cocoon as this can subsequently impact $t_{\rm br}$. For this work, we choose a time-step $\sim \mathcal{O}(0.1)$s.
\subsection{Effects due to time delay in launching of the jet ($t_{\rm lag}$)}
\label{subsec:res_tlag}
\begin{table*}
\centering
\def\arraystretch{1.5}
\caption{Table corresponding to Figure~\ref{fig:tbr_tlag}, shows whether for a given value of $L_{\rm j, iso}$ and $t_{\rm lag}$ the jet collimates and/or breaks out of the debris. We fix the value of expansion velocity of the debris $v_{\rm deb}=0.03c$ for all cases, as we vary the delay time for jet launching $t_{\rm lag}\sim10^{6-8}\,{\rm s}$. The notation used is: Collimation: the jet gets collimated prior to break out, No collimation: the jet does not get collimated before breaking out, Breakout: the jet successfully breaks out of the stellar debris, and No breakout: the jet gets choked with the debris material. The cases of collimated and choked jets in the table represent the scenario that would be ideal for neutrino production. Columns without entries imply that the entire row has the same outcome as stated in the middle column.}
\scalebox{0.95}{\begin{tabular}{ c c c c c }
\hline
$L_{\rm j,iso}$ (in ${\rm erg/s}$) & $t_{\rm lag} = 10^6\,{\rm s}$ & $t_{\rm lag} = 10^7\,{\rm s}$ & $t_{\rm lag} = 10^8\,{\rm s}$\\
\hline
$10^{40}$ -- $10^{43}$ & & Collimation; No breakout & \\
\hline
$5 \times 10^{43}$ & Collimation; No breakout & Collimation; No breakout & No Collimation; No breakout \\
\hline
$10^{44}$ & Collimation; No breakout & Collimation; No breakout & No collimation; No breakout \\
\hline
$5 \times 10^{44}$ & Collimation; Breakout & No collimation; No breakout & No collimation; No breakout \\
\hline
$10^{45}$ & Collimation; Breakout & No collimation; No breakout & No collimation; No breakout \\
\hline
$5 \times 10^{45}$ & Collimation; Breakout & No collimation; No breakout & No collimation; No breakout \\
\hline
$10^{46}$ & Collimation; Breakout & No collimation; Breakout & No collimation; No breakout \\
\hline
$5 \times 10^{46}$ & No collimation; Breakout & No collimation; Breakout  & No collimation; No breakout \\
\hline
$10^{47}$ & No collimation; Breakout & No collimation; Breakout & No collimation; No breakout \\
\hline 
\end{tabular}}
\label{tab:res1}
\end{table*}
The time delay associated with jet launching plays an important role in collimation and breakout of the jet. The origin of such a time delay is discussed in detail in Section~\ref{subsec:delay}. As this time delay in launching of the jet can range from $\sim$few days to a few years, we assume three different values of $t_{\rm lag}$: $\mathcal{O}(10\ \rm days) \sim 10^6\,{\rm s}$, $\mathcal{O}(100\ \rm days) \sim 10^7\,{\rm s}$, and $\mathcal{O}(1000\ \rm days) \sim 10^8\,{\rm s}$, which are denoted by red stars, purple unfilled squares and blue filled downward triangles, respectively, in Figure ~\ref{fig:tbr_tlag}. Note that for $t_{\rm lag} = 10^8$s, none of the jets in the examined range of $L_{\rm j,iso}$ break out so we do not see the corresponding contour in the figure. The results in terms of the breakout time $t_{\rm br}$ are shown for different values of $t_{\rm lag}$ as $L_{\rm j, iso}\sim10^{40-47}\,{\rm erg/s}$ is varied, and are also summarised in Table~\ref{tab:res1}, where we fix $v_{\rm deb} = 0.03c$.

We find collimated choked jets for the entire range of $t_{\rm lag}$ when $L_{\rm j,iso} \lesssim 10^{43}\ \rm erg/s$. As $t_{\rm lag}$ increases, the required energy for the jets to successfully breakout of the debris increases, and therefore, jets with higher luminosities breakout whereas the ones with lower luminosities get \emph{choked}. This can be understood from the fact that as the delay time for jet launching increases, the debris has more time to expand, making it harder for the jet to breakout. Also as expected, the jets with higher isotropic-equivalent luminosities break out much quicker than the ones with lower $L_{\rm j, iso}$. For $t_{\rm lag} = 10^8\,{\rm s}$, none of the jets breakout for the range of $L_{\rm j,iso}$ considered. This is reasonable since the time delay in jet-launching for this case is very large $\sim$ 1000 days, hence the jets mostly get choked. For a typical $t_{\rm lag} = 10^7\,{\rm s}$ ($\sim$ 100 days), we find jets with $L_{\rm j,iso} \lesssim 10^{44}\,{\rm erg/s}$ are collimated and choked.

We use equation~\eqref{eq:chokcrit} to match the analytical and numerical constraints for $L_{\rm j,iso}$ in the context of jet choking. From equation~\eqref{eq:chokcrit}, we find that for $v_{\rm deb} = 0.03c$ and the three values of $t_{\rm lag}$ considered i.e., $10^6\,{\rm s}$, $10^7\,{\rm s}$, and $10^8\,{\rm s}$, all jets with $L_{\rm j,iso}$ not exceeding $\sim 9.8 \times 10^{44}\,{\rm erg/s}$, $\sim 3.2 \times 10^{45}\,{\rm erg/s}$, and $\sim 9.8 \times 10^{46}\,{\rm erg/s}$, respectively, should get choked. These numbers approximately match with the ones found from our numerical simulations (see Figure~\ref{fig:tbr_tlag} and Table~\ref{tab:res1}). For lower values of $t_{\rm lag}$, the fact that the jet breaks out while being collimated introduces some small differences between the analytical and numerical results (as discussed in Section~\ref{subsec:jetbr}).

\subsection{Effects due to velocity of the wind ($v_{\rm deb}$)}
\begin{table*}
\centering
\caption{Same as Table~\ref{tab:res1} but for Figure~\ref{fig:tbr_vw}, for different values of the debris expansion velocity $v_{\rm deb}\sim 0.01c -0.1c$ with fixed $t_{\rm lag} = 10^7\,{\rm s}$.} 
\label{tab:res2}
\def\arraystretch{1.5}
\scalebox{0.95}{
\begin{tabular}{c c c c}
\hline
$L_{\rm j,iso}$ (in ${\rm erg/s}$) & $v_{\rm deb} = 0.01c$ & $v_{\rm deb} = 0.03c$ &  $v_{\rm deb} = 0.1c$\\
\hline
$10^{40}$ -- $10^{44}$ &  & Collimation; No breakout & \\
\hline
$5 \times 10^{44}$ -- $10^{45}$ & & No collimation; No breakout & \\
\hline
$5 \times 10^{45}$ & No collimation; Breakout & No collimation; No breakout & No collimation; No breakout \\
\hline
$10^{46}$ & No collimation; Breakout & No collimation; Breakout & No collimation; No breakout\\
\hline
$5 \times 10^{46}$ -- $10^{47}$ & & No collimation; Breakout &  \\
\hline
\end{tabular}
}
\end{table*}
The expansion velocity of the debris is another uncertain quantity that plays an important role in the outcome of the jet-cocoon dynamics. In general, $v_{\rm deb}$ is a function of both time and distance from the SMBH. In this work, as an approximation we take it to be constant, where we treat $v_{\rm deb} = 0.03c$ as the fiducial case and vary it between low ($v_{\rm deb} = 0.01 c$) and high ($v_{\rm deb} = 0.1 c$) velocity regimes. These can be thought of as the time-averaged values of $v_{\rm deb}$ for the duration of evolution of the jet-cocoon system. In Figure~\ref{fig:tbr_vw} and Table~\ref{tab:res2}, we show the results for the breakout time for the three values of $v_{\rm deb}$: $0.01$c as orange stars, $0.03$c as green unfilled squares, and $0.1$c as dark blue filled downward triangles, with varying $L_{\rm j,iso}\sim10^{40-47}\,{\rm erg/s}$. The effect of $v_{\rm deb}$ on collimating the jets with varying $L_{\rm j,iso}$ is however insignificant, since for all cases the jets get collimated for $L_{\rm j,iso} \lesssim 10^{44}$ergs/s and remain uncollimated for higher values of $L_{\rm j,iso}$.

In Figure~\ref{fig:tbr_vw}, for all values of $v_{\rm deb}$ considered, we note that $t_{\rm br}$ decreases as $L_{\rm j,iso}$ increases, which is expected since jets with higher luminosities are more energetic and hence breakout easily. For the jets with same $L_{\rm j,iso}$, the breakout time increases with increasing $v_{\rm deb}$. This is because for a higher $v_{\rm deb}$, $R_{\rm out}(t)$ expands more rapidly, and thus the jet requires more time to break out. This is also the reason why the limiting value of $L_{\rm j,iso}$ for choked jets increases as $v_{\rm deb}$ increases. For $v_{\rm deb} = 0.1c$ we see that most of the jets remain choked for the range of $L_{\rm j,iso}$ we consider and only the highest isotropic-equivalent luminosity jets ($L_{\rm j,iso} \gtrsim 5 \times 10^{46}$ erg/s) break out. This is because a higher debris velocity implies that the outer radius of the debris is significantly larger than the jet-head radius leading to a choked jet.

Once again, we can compare the analytical and numerical results using equation~\eqref{eq:chokcrit} to find the regime of choked jets for different values of $v_{\rm deb} \sim 0.01c - 0.1c$. We find analytically that for $t_{\rm lag} = 10^7\,{\rm s}$, $L_{\rm j,iso}$ should not exceed $\sim 3.6 \times 10^{44}\,{\rm erg/s}$, $\sim 3.2 \times 10^{45}\,{\rm erg/s}$, and $\sim 3.6 \times 10^{46}\,{\rm erg/s}$ for $v_{\rm deb} = 0.01c$, $0.03c$, and $0.1c$, respectively, for the jet to be choked by the wind-driven debris. As in the previous case, the analytical results for the critical luminosities approximately match with what we obtain numerically (see Figure~\ref{fig:tbr_vw} and Table~\ref{tab:res2}). 

\section{Implications for electromagnetic and neutrino counterparts}
\label{sec:signatures}
We studied the physical system pertaining to TDEs with accompanying jets and associated dynamics in the previous sections. Most importantly, we focused on jets that are launched with an associated delay of $t_{\rm lag}\sim10^{6-8}\,{\rm s}$ into a wind-driven debris expanding with velocity $v_{\rm deb}\sim0.01c-0.1c$. We also discussed the effects of $t_{\rm lag}$ and $v_{\rm deb}$ on the jet collimation and breakout. In this section, we discuss the observable multi-messenger signatures from \emph{delayed choked jets} in TDEs. Particle acceleration within jets can lead to the production of gamma rays, high-energy neutrinos as well as UHECRs (see e.g.,~\citealt{AlvesBatista:2017shr,Zhang:2017hom,Biehl:2017hnb,Guepin:2017abw,Bhattacharya:2021cjc,MB_2023}). We primarily focus on the electromagnetic signatures and qualitatively discuss the neutrino signatures.

\subsection{Electromagnetic signatures}
\label{sec:elemag}
The jet-head is initially relativistic but eventually slows down to sub-relativistic velocities due to its interaction with the debris and  the formation of the cocoon. The deceleration of such a relativistic outflow produces shocked regions in the jet-head. A forward shock propagates outwards and a reverse shock moves inwards towards the inner boundary of the debris. Relativistic electrons are accelerated at the shocks and they cool due to radiative losses through different processes such as synchrotron and inverse Compton emission. 
We focus on the synchrotron cooling process and resulting electromagnetic emission. The typical frequencies associated with synchrotron emission are: (a) the injection frequency $\nu_m$ corresponding to the injection Lorentz factor $\gamma_m$ of the accelerated electrons, (b) the cooling frequency $\nu_c$ of the electrons at which the radiative cooling timescale matches the expansion timescale, and (c) the absorption frequency $\nu_a$ associated with the synchrotron-self absorption (SSA) of the electrons which is relevant mainly at low frequencies. A generic expression of these fundamental break frequencies in the observer frame can be defined as,
\be
\label{eq:freqs}
\nu_\alpha^{\rm ES} = \frac{3}{4 \pi} \frac{e B^{\rm ES}}{m_e c} \frac{\Gamma^{\rm ES}}{(1+z)} \big(\gamma^{\rm ES}_\alpha\big)^2\,
\ee
where the subscript $\alpha=\{m,c\}$ corresponds to the injection and cooling frequencies, respectively. $\rm ES = \{\rm FS, RS\}$ corresponds to the forward and reverse shock regions, respectively. The magnetic field strength is $B^{\rm ES} = \big[ 32 \pi \epsilon_B \Gamma^{\rm ES} (\Gamma^{\rm ES} - 1) n^{\rm ES} m_p c^2 \big]^{1/2}$, where $n^{\rm ES}$ is the particle density and $\Gamma^{\rm ES}$ is the bulk Lorentz factor in the shocked region $\rm ES$. The fraction of downstream internal energy density that is converted to magnetic field energy density is given by $\epsilon_B$.

The minimum Lorentz factor of the electrons is defined as, $\gamma_m^{\rm ES} = \epsilon_e \zeta_e \big(\Gamma^{\rm ES} - 1\big)m_p/m_e$, where $\zeta_e \approx 1/\big[f_e (s-1)/(s-2) \big]$ for $s>2$ is constrained using particle-in-cell simulations~\citep[e.g.,][]{Park:2014lqa}, where the fraction of downstream internal energy density that is carried by non-thermal electrons is given by $\epsilon_e$. 
The power-law spectral index associated with the acceleration of electrons in the shocked region is given by $s$.
The fraction of accelerated electrons is given by $f_e$ and the maximum Lorentz factor associated with the electrons is defined as, $\gamma^{\rm ES}_M = \big( 6\pi e \big)^{1/2}/\big[ \sigma_T B^{\rm ES} (1+Y) \big]^{1/2}$, where $\sigma_T$ and $Y$ are the Thomson cross-section and Compton parameter, respectively. The maximum synchrotron frequency $\nu_M^{\rm ES}$ is given by equation~\eqref{eq:freqs}. 
The cooling Lorentz factor is defined as, $\gamma^{\rm ES}_c = 6 \pi m_e c/\big[ (1+Y) T^\prime \sigma_T (B^{\rm ES})^2 \big]$, where the comoving time $T^\prime$ is written in terms of the system evolution time $T$ and redshift $z$ as, $T^\prime = \Gamma^{\rm ES} T/(1+z)$. 
The SSA frequency $\nu_{sa}$ is obtained by setting $\tau_{\rm sa} (\nu=\nu_{\rm sa}) = 1$ and solving for $\nu_{sa}$, where the SSA optical depth as a function of the observed frequency $\nu$ is, $\tau^{\rm ES}_{\rm sa} (\nu) = \xi_s e n^{\rm ES} R_h (\nu/\nu^{\rm ES}_n)^{-p}/\big( B^{\rm ES} (\gamma^{\rm ES}_n)^5 \big)$ (see e.g., \citealt{Panaitescu:2000bk}). Here $p=3$ for the fast-cooling regime and is set to be $p=(4+s)/2$ for the slow-cooling regime.  
Using the above expression of $\tau_{\rm sa}$ and setting $\tau_{\rm sa} (\nu=\nu_{sa}) = 1$, we obtain, $\nu_{sa} = \nu^{\rm ES}_n \big[ \big( \xi_s e n^{\rm ES} R_h \big)/\big( B^{\rm ES} (\gamma^{\rm ES}_n)^5 \big) \big]^{1/p}$. The quantity $\xi_s \sim 5 - 10$ depends on the electron spectral index ($s$). We have the parameters $\gamma^{\rm ES}_n = \rm min \big[ \gamma^{\rm ES}_m, \gamma^{\rm ES}_c \big]$ and correspondingly $\nu^{\rm ES}_n = (\gamma^{\rm ES}_n)^2eB^{\rm ES}/\big( (1+z) m_e c \big)$. The peak synchrotron flux is given by, $F^{\rm ES}_{\rm syn, max} \approx 0.6 f_e n^{\rm ES} R_h^3 \Gamma^{\rm ES} e^3 B^{\rm ES} (1+z)/\big( \sqrt{3} m_e c^2 d_L^2 \big)$, where $d_L$ is the luminosity distance.

\begin{figure*}
\begin{center}
\subfloat[(a)] {\includegraphics[width=0.48\textwidth]{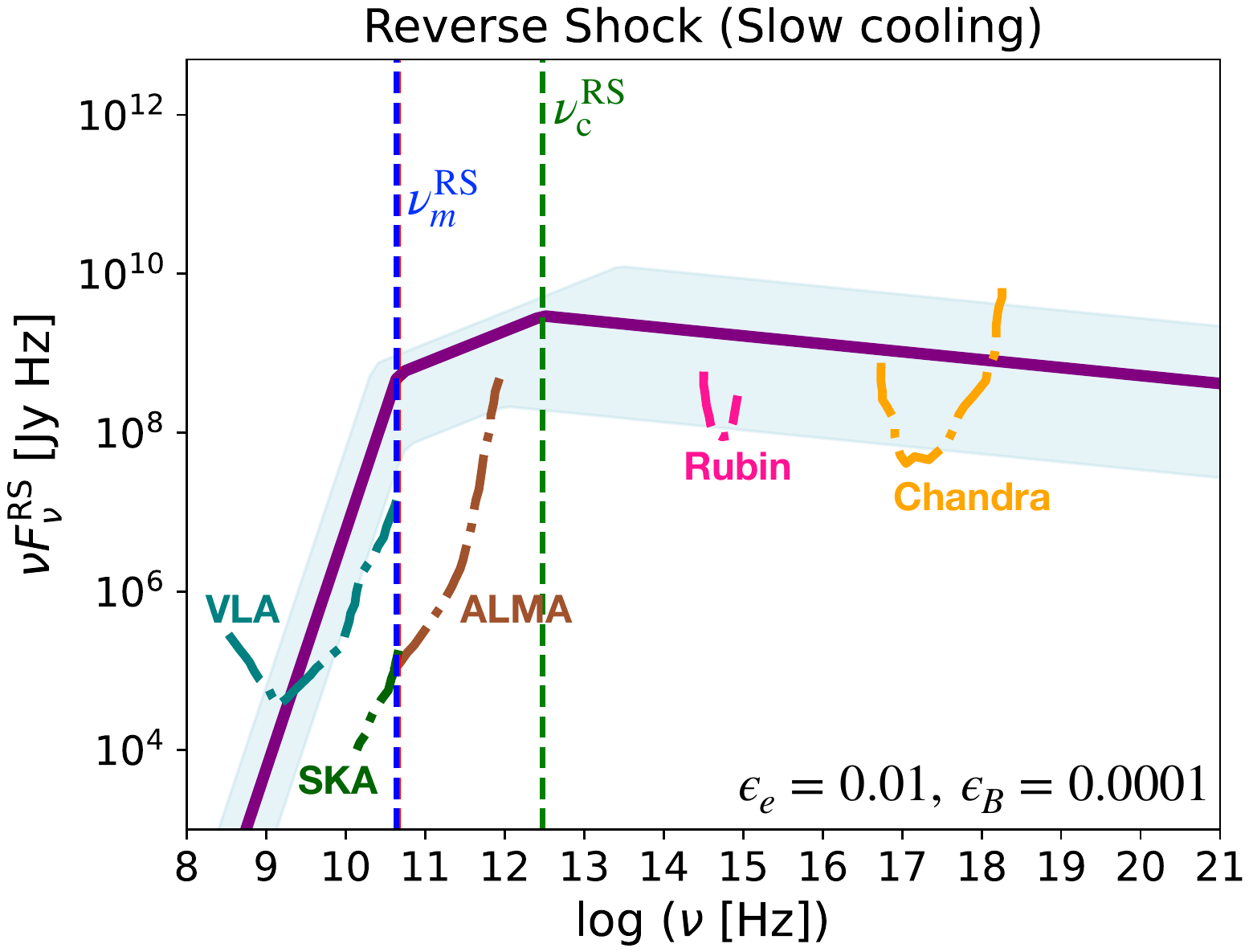}\label{fig:slowcool_rs}}\hfill
\subfloat[(b)] {\includegraphics[width=0.49\textwidth]{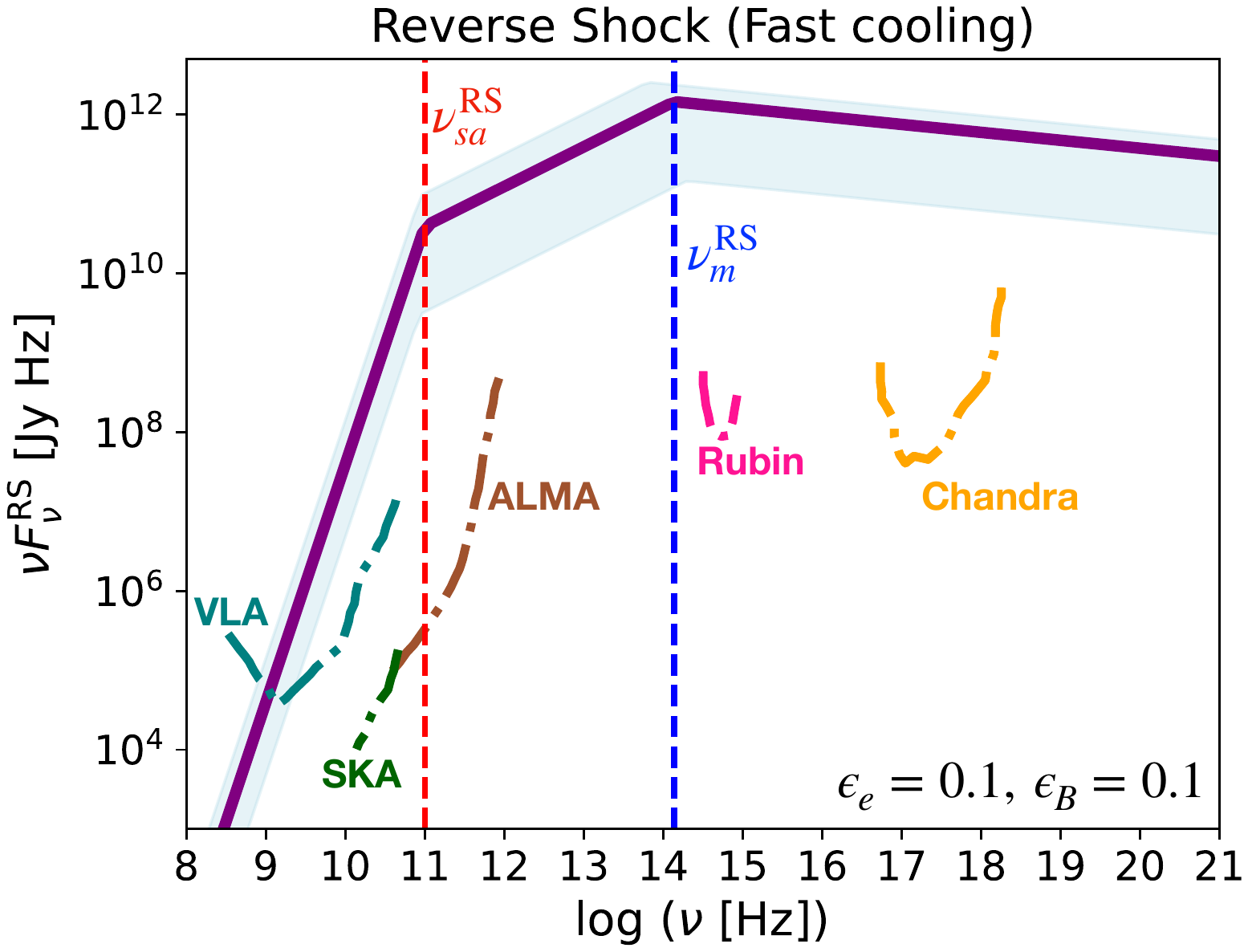}\label{fig:fastcool_rs}} \hfill
\caption{\label{fig:em_rs} Electromagnetic signatures from the reverse shock (RS) emission region for the maximum isotropic-equivalent luminosity possible from a delayed choked jet scenario are shown for a TDE source located at redshift $z=0.05$ and assuming $v_{\rm deb} = 0.03c$ (solid purple curve): \emph{Left:} spectral energy density for synchrotron slow cooling regime with $\epsilon_e = 0.01$ and $\epsilon_B = 0.0001$. \emph{Right:} spectral energy density for synchrotron fast cooling regime with $\epsilon_e = 0.1$ and $\epsilon_B = 0.1$. The dashed vertical lines show the characteristic emission frequencies ($\nu_a^{\rm RS}$, $\nu_m^{\rm RS}$, $\nu_c^{\rm RS}$) for both cases (for the left panel $\nu_a^{\rm RS}$ approximately coincides with $\nu_m^{\rm RS}$ and hence is not visible, whereas for the right panel $\nu_c^{\rm RS}$ lies outside the range of frequencies shown and hence is not visible). The upper and lower boundaries of the shaded region show the corresponding spectral energy density for $v_{\rm deb} = 0.1c$ and $v_{\rm deb} = 0.01c$, respectively. The dot-dashed lines show the sensitivity curves for various detectors in X-ray (Chandra - $100$ ks), optical (Vera C. Rubin Observatory - LSST - $30$ s) and radio (ALMA - $1$ hr, SKA - $10$ hr, VLA - $1$ hr) bands. 
The relevant information regarding these facilities can be found at: Chandra (\href{https://chandra.harvard.edu}{https://chandra.harvard.edu}), Vera C. Rubin Observatory - LSST (\href{https://www.lsst.org}{https://www.lsst.org}), ALMA (\href{https://alma-telescope.jp/en/}{https://alma-telescope.jp/en/}), SKA (\href{https://www.skao.int/en}{https://www.skao.int/en}), and VLA (\href{https://public.nrao.edu/telescopes/vla/}{https://public.nrao.edu/telescopes/vla/}).
}
\end{center}
\end{figure*}

The electron cooling follows the fast or slow cooling regime depending on the values of $\nu^{\rm ES}_m$ and $\nu^{\rm ES}_c$. When the dynamical timescale of the system is much longer than the timescales over which the electrons cool down due to radiation losses, we have $\nu^{\rm ES}_m>\nu^{\rm ES}_c$ which is known as the \emph{fast cooling} regime. In this case, all the electrons rapidly cool down to $\gamma_e \sim \gamma_c$. When $\nu^{\rm ES}_c > \nu^{\rm ES}_m$, only electrons with Lorentz factors greater than $\gamma^{\rm ES}_c$ can cool and this regime is known as the \emph{slow cooling} regime. We will now discuss the electromagnetic signatures from choked jets, from both forward and reverse shock regions. For this analysis, we assume $s=2.2$ and $\zeta_e = 0.35$, which implies $\xi_s \sim 7.5$. The TDE event is assumed to be located at a redshift $z = 0.05$, which is also similar to AT2019dsg.
\subsubsection{Forward shock}
The forward shock moves outwards with the jet-head similar to a blast wave. In this region we have $\rm ES = \rm FS$, the associated bulk Lorentz factor $\Gamma^{\rm FS} = \Gamma_h$, and the particle number density $n^{\rm FS} = \rho_a/m_p$ (see Section~\ref{sec:model}). As an example, we choose the highest isotropic-equivalent luminosity ($L_{\rm j,iso} = 5 \times 10^{45}\,{\rm erg/s}$) for which we get a choked jet, with an associated delay $t_{\rm lag} = 10^7\,{\rm s}$ and $v_{\rm deb} = 0.03c$. Substituting $\epsilon_B = 0.0001$ in equation~(\ref{eq:freqs}) and the above equations, we calculate the characteristic frequencies for this case. We find that $\nu^{\rm FS}_m < \nu^{\rm FS}_c < \nu^{\rm FS}_{sa}$, which implies that we are in the slow cooling regime. However, in this case the absorption frequency is the highest and hence most of the emission would appear around $\nu_{sa}$~(see e.g.,~\citealt{Kobayashi:2003yh}).
\subsubsection{Reverse shock}
Reverse shock can be accompanied by the outward moving forward shock and can also be important for electromagnetic emissions. Here we focus on the emission associated with the reverse shock, which is analogous to the emission from hot spots of radio galaxies~\citep{Tingay:2008sg,Mack:2008wf}. In this case, we have $\rm ES= \rm RS$, the associated bulk Lorentz factor $\Gamma^{\rm RS} = 0.5 \big( \Gamma_0/\Gamma^{\rm FS} + \Gamma^{\rm FS}/\Gamma_0 \big)$, where $\Gamma^{\rm FS} = \Gamma_{\rm h}$ and we assume $\Gamma_0 = 10$. The particle number density associated with the reverse shock region is  $n^{\rm RS} = L_{\rm j,iso}/\big( 4 \pi R_h^2 \Gamma_0^2 m_p c^3 \big)$. To explore the different scenarios of slow and fast cooling that can be associated with the reverse shock region, we consider two different cases: one with typical values of $\epsilon_B=0.0001$ and $\epsilon_e=0.01$, and another where $\epsilon_B=0.1$ and $\epsilon_e=0.1$, as discussed below.

A typical case can be considered where we assume $\epsilon_B = 0.0001$ and $\epsilon_e = 0.01$. We find the characteristic frequencies to be ordered as $\nu^{\rm RS}_c > \nu^{\rm RS}_{m} > \nu^{\rm RS}_{sa}$ from equation~\eqref{eq:freqs}, and therefore, the \emph{slow cooling regime} is relevant for this case. The characteristic frequencies are shown as dashed vertical lines in Fig.~\ref{fig:slowcool_rs} and are given as $\nu^{\rm RS}_c \simeq 3.0 \times 10^{12}\,{\rm Hz}$, $\nu^{\rm RS}_{\rm sa} \simeq 4.5\times 10^{10}\,{\rm Hz}$, and $\nu^{\rm RS}_m \simeq 4.4 \times 10^{10}\,{\rm Hz}$ in green, red, and blue respectively. Note that since $\nu_{sa}^{\rm RS} \sim \nu_m^{\rm RS}$, in the figure the two lines overlap and the red dashed line corresponding $\nu_{sa}^{\rm RS}$ is not visible). We can estimate the peak synchrotron flux as, $F_{\rm syn,max}^{\rm RS} \simeq 37\ {\rm mJy}\ (f_e/0.48) n^{\rm RS}_{2.53} R^3_{\rm h,16.21} \Gamma^{\rm RS}_{0.70} B^{\rm RS} (1+z) d_{L,26.82}^{-2}$, where we assume $s=2.2$, $f_e = 1/\big[\zeta_e (s-1)/(s-2) \big]$ and $B^{\rm RS} = \big[ 32 \pi \epsilon_B \Gamma^{\rm RS} (\Gamma^{\rm RS} - 1) n^{\rm RS} m_p c^2 \big]^{1/2} \simeq 0.32\,{\rm G}$. To illustrate the plausible electromagnetic emission spectrum, we show the results for a choked delayed jet with the highest $L_{\rm j,iso} = 5 \times 10^{45}\,{\rm erg/s}$ corresponding to $t_{\rm lag} = 10^7\,{\rm s}$, and $v_{\rm deb} = 0.03c$ as considered for the forward shock. In this case, given the ordering of the characteristic frequencies, the observed spectrum $F^{\rm RS}_\nu$ is,
\be
\label{eq:slowcool}
F^{\rm RS}_\nu = F^{\rm RS}_{\rm syn, max}
\hspace{-0.75mm}\left\{ \hspace{-2.75mm}
\begin{array}{ll}
\big( \nu^{\rm RS}_{m}/\nu^{\rm ES}_{sa} \big)^{(s+4)/2} \big( \nu/\nu^{\rm RS}_m \big)^{2}, & \hspace{-3mm} \nu \leq \nu^{\rm RS}_{m} \vspace{0.3cm} \\
\big( \nu^{\rm RS}_{sa}/\nu^{\rm RS}_m \big)^{-\frac{(s-1)}{2}} \big( \nu/\nu^{\rm RS}_{sa} \big)^{-\frac{5}{2}}, & \hspace{-3mm} \nu^{\rm RS}_{m} <\nu \leq \nu^{\rm RS}_{sa} \vspace{0.3cm} \\
\big( \nu/\nu^{\rm RS}_m \big)^{-(s-1)/2}, & \hspace{-3mm} \nu^{\rm RS}_{sa} <\nu \leq \nu^{\rm RS}_c \vspace{0.3cm} \\
\big( \nu^{\rm RS}_c/\nu^{\rm RS}_m \big)^{-\frac{(s-1)}{2}} \big( \nu/\nu^{\rm RS}_c \big)^{-\frac{s}{2}}, & \hspace{-3mm} \nu^{\rm RS}_c <\nu \leq \nu^{\rm RS}_M
\end{array}
\right.
\ee
The synchrotron slow cooling spectrum of electromagnetic emissions from equation~\eqref{eq:slowcool} is shown with a solid purple curve in Figure~\ref{fig:slowcool_rs}.

As expected, the spectrum peaks at the cooling frequency $\nu^{\rm RS}_c$. The electromagnetic observability of the synchrotron emission would mostly be in the optical and X-ray bands, where a current X-ray telescope like Chandra and/or an optical telescope like Vera C. Rubin Observatory (LSST) have the possibility to definitely detect the electromagnetic signals from a nearby TDE like AT2019dsg. Some radio detections also seem promising with telescopes such as ALMA (Atacama Large Millimeter/submillimeter Array) and SKA (Square Kilometre Array). But the spectrum rapidly declines due to the sharp cut-off for the synchrotron spectrum in the radio band. 
We also consider the cases for $v_{\rm deb} = 0.01c$ and $0.1c$, as the lower and upper boundaries of the shaded regions in both figure panels. 

When $\epsilon_B = 0.1$ and $\epsilon_e = 0.1$, we can calculate the characteristic frequencies from equation~(\ref{eq:freqs}) to find $\nu^{\rm RS}_m > \nu^{\rm RS}_{sa} > \nu^{\rm RS}_{c}$, which implies that the synchrotron emission occurs in the fast cooling regime. In this case, the characteristic frequencies $\nu^{\rm RS}_m$ (dashed blue line), $\nu^{\rm RS}_c$ (dashed green line) and $\nu^{\rm RS}_{sa}$ (dashed red line) are $1.4 \times 10^{14}\,{\rm Hz}$, $9.6 \times 10^{7}\,{\rm Hz}$ and $1.0 \times 10^{11}\,{\rm Hz}$, respectively, as shown in Fig.~\ref{fig:fastcool_rs}. Note that the value of $\nu_c^{\rm RS}$ is such that it does not fall in the range of frequencies shown in the figure, so we do not see the dashed green line. The peak synchrotron flux for this case can be estimated as, $F_{\rm syn,max}^{\rm RS} \simeq 37\ {\rm mJy}\ (f_e/0.48) n^{\rm RS}_{2.53} R^3_{\rm h,16.21} \Gamma^{\rm RS}_{0.70} B^{\rm RS} (1+z) d_{L,26.82}^{-2}$, where $B^{\rm RS} \simeq 10.25\,{\rm G}$. The observed spectrum for the fast cooling case is given by,
\be
\label{eq:fastcool}
F^{\rm RS}_\nu = F^{\rm RS}_{\rm syn, max}
\hspace{-0.75mm} \left\{ \hspace{-2.75mm}
\begin{array}{ll}
\big( \nu/\nu^{\rm RS}_{sa} \big)^{2}, & \hspace{-3mm} \nu \leq \nu^{\rm RS}_{sa} \vspace{0.3cm} \\
\big( \nu/\nu^{\rm RS}_{\rm sa} \big)^{-1/2}, & \hspace{-3mm} \nu^{\rm RS}_{sa} <\nu \leq \nu^{\rm RS}_m \vspace{0.3cm} \\
\big( \nu^{\rm RS}_m/\nu^{\rm RS}_{\rm sa} \big)^{-1/2} \big( \nu/\nu^{\rm RS}_m \big)^{-s/2}, & \hspace{-3mm} \nu^{\rm RS}_m <\nu \leq \nu^{\rm RS}_M
\end{array}
\right.
\ee
For this regime, we compute the electromagnetic emission spectrum for $L_{\rm j,iso}= 5 \times 10^{45}\,{\rm erg/s}$ and $t_{\rm lag}=10^7\,{\rm s}$, and $v_{\rm deb} = 0.03c$ using equation~(\ref{eq:fastcool}), and the corresponding results are shown in Figure~\ref{fig:fastcool_rs}. In this case, the spectrum peaks at $\nu^{\rm RS}_m$. We see that the spectrum is roughly two orders of magnitude higher than that of the slow cooling case. However, the choices of $\epsilon_B$ and $\epsilon_e$ are optimistic and may not be strictly true. Besides, when $\epsilon_B \sim \epsilon_e$, the inverse Compton process might be equally important as synchrotron. 

Keeping these assumptions in mind, we note observationally that this is a more optimistic case and the electromagnetic signatures can definitely be detected with the current generation X-ray and optical telescopes upto $z \sim 1$. However, radio observation prospects are very similar to the previous slow-cooling scenario, where ALMA and SKA can be good candidates to observe the radio emission from nearby ($z \sim 0.05$) TDEs. In this case as well, we show the cases of $v_{\rm deb} = 0.01c$ and $0.1c$ as the lower and upper boundaries of the shaded region.
It is important to discuss the timescales over which the reverse shock emission happens such that detection by telescopes can be well-timed and have maximum efficiency. This timescale is given by $(R_{\rm h}-R_{\rm in})/\Gamma_{\rm h}^2 c$, which is the observation time as the jet  propagates through the debris. Since we are interested in choked jets and discuss the emission at the final instant when the jet gets choked, for us this is roughly $\sim 10^7\,{\rm s}$ after the TDE occurs. Thus electromagnetic observations would be prominent around $\gtrsim 100$ days after the TDE.

\subsubsection{Interaction between jet-driven debris and circumnuclear material}
In the previous section, we looked at the electromagnetic signatures from delayed choked jets, and showed that radio observations of the reverse shock emission may be promising for a nearby TDE source that is associated with delayed choked jet. However, we note that after the jet gets choked a small fraction of the cocoon surrounding the jet-head and debris are accelerated, which can then interact with the surrounding circumnuclear matter to produce radio emission. Given the low density of the cirucumnuclear matter, this emission could be visible with the radio telescopes.

A detailed analysis of such late-time radio followup observations associated with TDEs was performed in~\cite{Generozov:2016oon,Matsumoto:2021gcg}, where 
disk winds, unbound debris, conical outflows and relativistic jets were considered. However, for this work we consider $\eta_{\rm fb} \sim 0.5$ which determines the maximum fraction of stellar debris that can lead to the formation of a bound envelope. Hence, the number density of the electrons at the time of jet choking is still high for the case we discussed above. Therefore, observable radio emission can be expected in this case from the subsequent interaction of the jet with the circumnuclear material.

\subsection{High-energy neutrino signatures}
\label{sec:nu}
Neutrino emission from TDEs has been extensively discussed especially after the detection of two candidate neutrino-emitting TDEs, AT2019dsg~\citep{Stein:2019ivm} and AT2019fdr~\citep{Reusch:2021ztx}. 
Successful jets are disfavored due to the lack of sufficient observational evidence, including the absence of relativistic afterglows~\citep{Murase:2020lnu,Cendes:2021bvp,Matsumoto:2021qqo}. Choked TDE jets may explain the lack of electromagnetic counterparts from these relativistic outflows, but are also unlikely to be sufficiently powerful to explain the observed neutrino emission \citep{Murase:2020lnu}. If the jet is delayed, in principle, more powerful jets can also be choked within the expanding debris and we discuss the implications for their high-energy neutrino observations here.   

\begin{figure}
\includegraphics[width=0.49\textwidth]{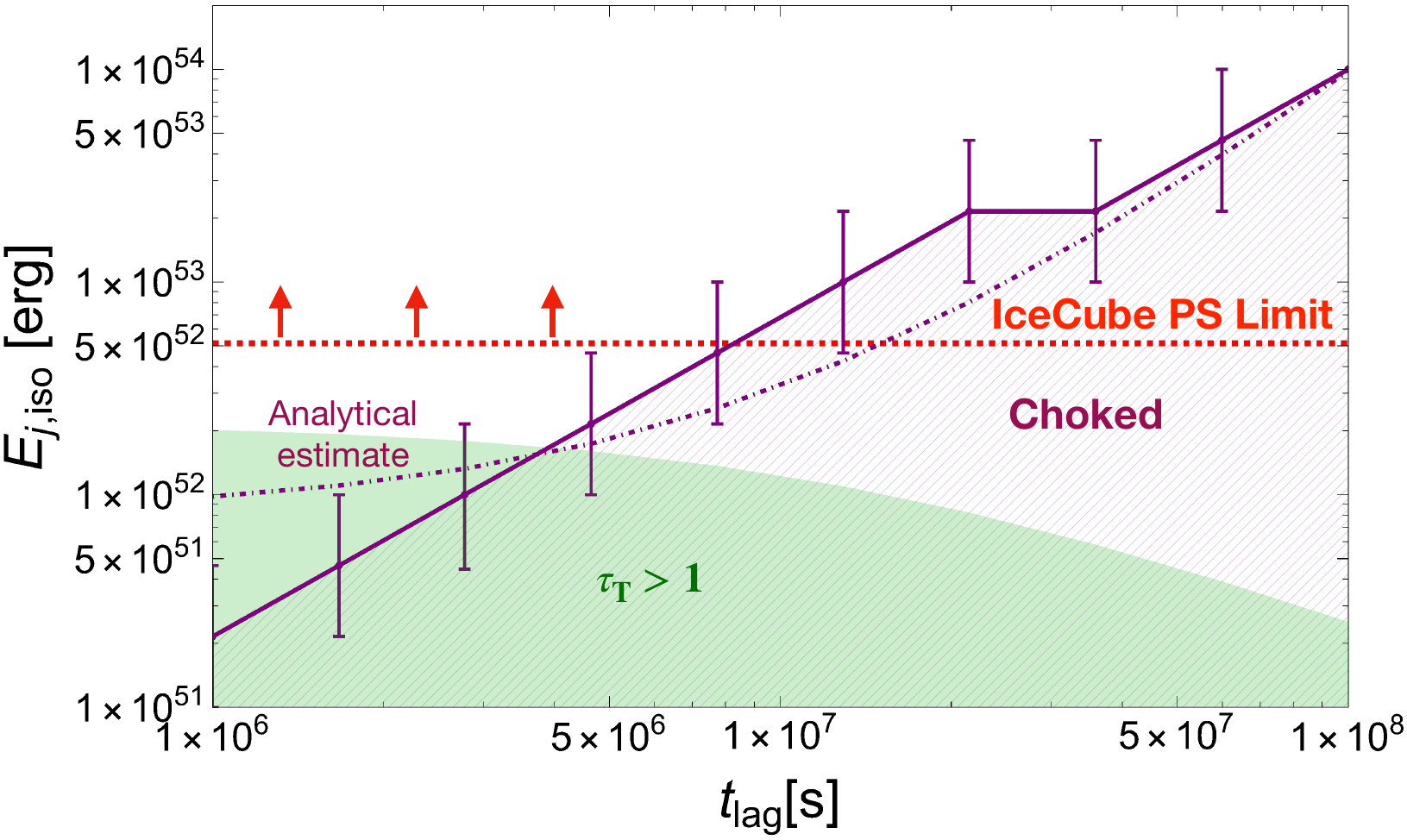}
\vspace{-0.3cm}
\caption{\label{fig:nuulplot} Figure showing the boundary (solid purple curve) for the isotropic jet energy, $E_{\rm j,iso}$, below which the jet is necessarily choked (hatched purple lines) for different values of associated $t_{\rm lag}$. Each numerical data point also has a corresponding error bar based on the discrete scanning of the range of luminosities. The IceCube limit for point sources (PS) (dashed red line) required to produce one neutrino event in the PS channel assuming a $E_\nu^{-2}$ spectrum is also shown, where the limit has been adapted to be plotted as energy (see the text for details). The analytical estimate of associated critical luminosity for choked jets is also plotted as a dot-dashed purple curve using equation~\eqref{eq:chokcrit}. The green shaded area shows the region where the Thomson optical depth $\tau_{\rm T} > 1$ in the shocked region. The source is assumed to be at a redshift of $z=0.05$ and with $v_{\rm deb} = 0.03c$.
}
\end{figure}

In Figure~\ref{fig:nuulplot}, we show the boundary between the successful and choked jets for different values of $t_{\rm lag}\sim10^{6-8}\,{\rm s}$, plotted with the isotropic energy $E_{\rm j, iso} = L_{\rm j,iso}t_{\rm dur}$ associated with the jet. $t_{\rm dur} = 10^7\,{\rm s}$ is used and the debris velocity is fixed at the fiducial value of $v_{\rm deb} = 0.03c$. 
The hatched orange region shows the values of $E_{\rm j,iso}$ for which choked jets are realized. As expected, jets with larger $E_{\rm j, iso}$ are more likely to get choked for a larger value of $t_{\rm lag}$ as the debris expands further. For each numerical data point shown in the figure we also show a corresponding error bar based on the intervals assumed in scanning through the range of $L_{\rm j,iso}$. The last data point does not have an error bar since we restrict $L_{\rm j,iso} \lesssim 10^{47}\,{\rm erg/s}$ and do not scan for higher values. We also show the analytical curve obtained using equation~\eqref{eq:chokcrit} as a dot-dashed purple line. In the expanding debris, we define the Thomson optical depth as $\tau_{\rm T}(t) = \int_{R_{\rm h}(t)}^{R_{\rm out}(t)} \kappa_{\rm es}\ \rho_{\rm deb}(t,r) dr$, where $\kappa_{\rm es}=0.35\, {\rm cm^2/g}$ is the Rosseland mean opacity. The region of $E_{\rm j, iso}$ where we obtain $\tau_{\rm T} > 1$ is shaded in green. 

The neutrino fluence necessary to produce a single event in the point source (PS) channel is $\sim 0.05\,{\rm GeV\,cm^{-2}}$ for a $E_\nu^{-2}$ spectrum, based on the public effective area~\citep{IceCube:2019cia,Murase:2020lnu}. The PS effective area leads to more conservative limits on the total energy than the Gamma-ray Follow Up (GFU) effective area. 
We convert the PS fluence limit to $E_{\rm j, iso}$ assuming a source redshift of $z=0.05$. We also assume that $\sim1$\% of the isotropic jet energy is used for neutrino production to get the dashed red line shown in Figure~\ref{fig:nuulplot}. This approximation is based on $E_{\nu_\mu} \approx (1/8) E_{\rm CR}/\mathcal{R}_{\rm CR}$, where the pre-factor of $1/8$ is due to the fact that same number of charged and neutral pions are obtained through the $p\gamma$ interactions from direct production, thereby leading to a factor of $1/2$. Furthermore, each neutrino flavor carries a quarter of the pion's energy in the decay chain post mixing. Here, $E_{\rm CR}$ is the total cosmic ray (CR) energy and ${\mathcal R}_{\rm CR} \sim 10-30$ is a bolometric correction factor associated with a CR spectral index of $s_p=2$. We also assume $E_{\rm CR, iso}=E_{\rm j, iso}$ to obtain the most conservative limit on the jet energy.
We note that the jets launched with an associated delay of $t_{\rm lag} \gtrsim 8\times10^6\,{\rm s}$ can lead to neutrino production above the current IceCube PS limit at this particular source distance.

Figure~\ref{fig:nuulplot} represents a scenario similar to the detection of IceCube-191001A with AT2019dsg at $z \sim 0.05$, and we note that our model of delayed choked jets can satisfy the lower limits with reasonable choice of parameters. 
We have also checked the case of the other two TDEs.  
The TDE AT2019aalc associated with IceCube-191119A was located at $z=0.036$ with a declination for which a similar PS effective area can be used. For this case, the nearby distance results in a similar value of $t_{\rm lag}$ to explain the associated neutrino event. The third event IceCube-200530A associated with AT2019fdr was localised at $z=0.267$. Using a similar PS effective area owing to a similar source declination, one obtains the energy required by the jet to produce a single neutrino event to be $\sim 2 \times 10^{54}\,{\rm erg}$ which shifts the horizontal line even further up. However, this scenario requires a time delay of $t_{\rm lag} \gtrsim  10^8\,{\rm s}$ which is longer than the observed delay. Thus we conclude that although the neutrinos from the two nearby TDEs (AT2019dsg and AT2019aalc) can be consistent with the delayed choked jet scenario, explaining the neutrino observation associated with AT2019fdr is still challenging with reasonable assumptions and choice of parameters.

We focus on choked TDE jets that can arise if the outflow is not sufficiently powerful to break out of the stellar envelope. 
We follow the prescription discussed in \cite{Senno:2016bso} to qualitatively discuss neutrino production in these choked jets. 
Particles can get accelerated in the internal shock (IS) site inside the collimation shock or termination shock close to the jet-head. We consider the IS scenario which occurs due to collisions between mass shells propagating with different velocities within the outflow. The energy dissipation takes place at the radius $R_{\rm IS} \approx \Gamma_j^2 c\delta t/(1+z) 
\sim (3\times10^{14}\,{\rm cm})(\delta t/100\ {\rm s}){(\Gamma_j/10)}^2$, where $\Gamma_j$ is the initial jet Lorentz factor and $\delta t$ is the temporal variability, as inferred by the X-ray emission from the jetted TDE Swift J1644+57~\citep{Burrows:2011dn}. For TDE jets, both internal and termination shocks are collisionless and radiation unmediated, which results in efficient particle acceleration up to ultrahigh energies (see e.g.,~\citealt{Farrar:2008ex}).  

Accelerated protons can generate neutrinos through both $pp$ and $p\gamma$ interactions. However, in the hidden (choked) jet model we consider, a much larger number density of photons implies $t_{p\gamma}^{-1} \gg t_{pp}^{-1}$, and therefore, the $p\gamma$ process is the primary neutrino production mechanism. 
In the IS region, the seed photons of $p\gamma$ interactions are expected to be non-thermal within the jet. If $\tau_T\gg1$, only a fraction $f_{\rm esc} \sim 1/\tau_T$ of photons can escape from the optically-thick envelope to the optically-thin jet region. We first estimate the fraction of protons, $f_{p\gamma}\approx t_{\rm dyn}/t_{p\gamma}$, which get accelerated in these choked TDE jets to undergo $p\gamma$ interactions and produce neutrinos. For a photon spectrum $n_{\varepsilon'} \propto {\varepsilon'}^{-\beta}$, the efficiency is roughly $f_{p\gamma} \propto {\varepsilon'}_p^{\beta-1}$, where ${\varepsilon}'_p$ is the proton energy in the jet-comoving frame. Using this $f_{p\gamma}$ estimate, one can compute the neutrino flux from a single TDE event assuming a flat energy spectrum ($\varepsilon_p L_{\varepsilon_p} \propto {\rm const}$), although the spectrum of high-energy neutrinos is modified when the pions and muons cool prior to their decay.

We also find that low-luminosity jets ($L_{\rm j,iso}\sim 10^{40} - 10^{43}\,{\rm erg/s}$) are collimated and choked, irrespective of $t_{\rm lag}$ and $v_{\rm deb}$, and hence can be ideal neutrino production sites. However, the energy budget available for neutrino production is also low owing to the smaller jet luminosities. For $t_{\rm lag} \sim 10^7\,{\rm s}$, we find a luminosity range $L_{\rm j,iso}\sim 10^{43} - 5 \times 10^{45}\,{\rm erg/s}$ where the jet gets choked.

\section{Summary and Discussion}
\label{sec:disc}
Recent observations of TDEs have indicated delayed or long-term activities that are seen much later than the peak time in optical emission. The origin of such delayed emissions from TDEs is not well understood, but it is likely to be powered by outflows launched from the central SMBH. In this work, we primarily focused on a scenario where the jet is launched with a delay of a few days to a few years. We studied the effects of this time lag on the jet dynamics and its interaction with the debris. We also discussed multi-messenger emission signatures expected in this scenario, in particular when the jet is choked. The physical model we consider involves an expanding debris that is presumably powered by the wind. Once a cocoon is formed due to the jet-debris interaction, the jet can get collimated depending on its isotropic-equivalent luminosity and the debris expansion velocity. Eventually, the jet either breaks out leading to a successful jet or gets choked within the debris.

We focused on the impacts of the time delay $t_{\rm lag}$ associated with jet launching and the debris expansion velocity $v_{\rm deb}$. We examined the effects of varying these parameters on the jet collimation and breakout conditions (see Figure~\ref{fig:res_tlag}). The critical luminosity for choked jets as a function of these parameters is obtained in equation~\eqref{eq:chokcrit}. These results are in agreement with the previous analytical estimate~\citep{Murase:2020lnu} in the limit of $t_{\rm lag}=0$.  Similarly, the critical choked jet luminosity found for delayed jets in binary neutron star mergers~\citep{Kimura:2018vvz}  are also in agreement with the analytical results obtained here. We found that for a longer $t_{\rm lag}$, it is easier for the jets to get choked within the debris. The corresponding $t_{\rm br}$ also increases for successful jets with a given luminosity. Similarly, a larger $v_{\rm deb}$ also makes the jets choked more easily and increases $t_{\rm br}$ for successful jets with a given luminosity.

The model presented in this work relies on several assumptions and approximations. One of the uncertainties is the timescale of the envelope formation. The fraction of disrupted stellar material that falls back onto the SMBH is assumed to be $0.5$, which can have related uncertainties. It is important to note that although the initial conditions are subject to different uncertainties, our main conclusions in this work are largely insensitive to them. The expression used for the effective tidal disruption radius $R_{\rm T}$ contains a factor $f_T$ associated with the shape of the stellar density profile. This factor can have uncertainties associated with it. For example,~\cite{Golightly:2019jib} considered the case of rotating stellar progenitors, in which case $R_{\rm T}$ is modified as, $R_{\rm T} = R_{\rm tidal} \big( 1 \pm \Omega_*^2R_*^3/(G M_*) \big)^{-1/3}$, where $\Omega_*$ is the spin of the progenitor star. Moreover, in~\cite{Ryu:2020huz}, the authors considered a fully general relativistic simulation of TDEs for stellar masses ranging from $0.15 M_\odot$ to $10 M_\odot$. It was concluded that the tidal disruption radius for low mass stars $M_* \leq 0.7 M_\odot$ is larger by $\sim 10$\%, whereas for $M_* \geq 1 M_\odot$ it is smaller by a factor of $\sim 2 - 2.5$. The value of $f_T$ mildly affects the effective tidal disruption radius $R_T \propto f_T^{1/6}$, since the dependence on $f_T$ is weak. However, for late times this effect is not prominent which can be seen from the approximate analytical expression for choking (equation~\ref{eq:chokcrit}). Thus, for the results related to choking or breakout of the jet and the corresponding EM and neutrino emissions, the effects of $f_T$ on the effective tidal disruption radius can be ignored.

As discussed in Section~\ref{subsec:jetpropej}, the calibration factor used for $\tilde{L}$ is largely unconstrained for TDE jets, primarily due to the lack of numerical simulations. However, one can reasonably assume it to be similar to the collapsar or binary merger systems. The cocoon pressure is assumed to be a simplified version of equation~(\ref{eq:pc}), where the integral is replaced by an average value (see equation~\ref{eq:avgint}). We verified that this approximation does not affect our results in any significant manner. The debris setup used in this work is rather simplified, and a smaller debris mass in the direction of jet propagation will result in the jet breaking out more readily, and therefore, the allowed power of choked jets is optimistic. A similar conclusion for the hidden/choked jet model is obtained by \citet{Murase:2020lnu}. However, we stress that in our setup the debris has longer time to expand for $t_{\rm lag} \sim 10^6 - 10^8\,{\rm s}$, allowing the choked jets to have larger luminosities (see Table~\ref{tab:res1}) and thereby larger total energies. Thus, we effectively relaxed the critical luminosity estimate found in \citet{Murase:2020lnu} using the current model.

We studied electromagnetic and neutrino signatures of choked TDE jets. For the electromagnetic emission, we limited ourselves to considering the synchrotron spectrum in the context of both forward and reverse shock scenarios. Specifically, we considered the emission from a choked jet with its maximum allowed luminosity $L_{\rm j,iso} = 5 \times 10^{45}\,{\rm erg/s}$ for an associated time delay of $t_{\rm lag} = 10^7\,{\rm s}$ and $v_{
\rm deb} = 0.03c$. We found that the forward shock emission may be largely suppressed due to strong synchrotron self-absorption, while the reverse shock emission, which is analogous to emission from hot spots of radio galaxies, is promising. Electromagnetic signatures of choked jets from nearby TDEs such as AT2019dsg at $z = 0.05$ may be observed with radio, X-ray and optical telescopes. Once the jet gets choked, a small part of the debris may be accelerated, leading to subsequent interactions with the circumnuclear material. This may also result in observable radio signatures. 

Successful observations of electromagnetic signatures will provide significant implications for the dynamics of TDEs, in particular the existence of a jet and a delayed launching scenario. 
It would also show evidence of the jet interacting with the debris. In general, the debris does not have a spherical configuration\footnote{Note that the relevant density profile for the ambient medium should be regarded as the density profile along the jet direction and therefore the geometry of the debris is not relevant strictly speaking.} \citep{Lu:2019hwv}, which leads to uncertainty in modeling the jet-debris interaction, and we will obtain more insights into the properties and mechanism of the disk formation with future observations.
The absence of such electromagnetic signatures from nearby TDE sources might further constrain the existence of jetted TDEs. It may also hint towards the fact that the jet-debris interaction is inefficient and hence it does not produce a detectable electromagnetic signature. 
Hydrodynamic simulations will also help to better understand the launch and formation of delayed jets and it would be useful to perform simulations of the jet propagating through an outward expanding debris aided by wind outflows. 

The delayed arrival of neutrinos could be associated with not only the disk transition~\citep{Murase:2020lnu} but also the formation of the inner accretion disk post circularization of the debris which takes $\sim 10^7\,{\rm s}$ based on analytical estimates~\citep{vanVelzen:2021zsm, Hayasaki:2021ggg}, assuming that the debris circularizes through dissipation at the self-interaction shock. A similar circularization timescale is obtained from the analytical estimates of~\cite{Hayasaki:2019kjy}, who assume that the dissipated energy is proportional to the mass fallback rate once the debris circularizes. In this case, the energy dissipated during the debris circularization is given by the product of the stellar mass and the difference between the specific stellar orbital energy and the specific binding energy of the debris at the circularization radius. Detailed hydrodynamic simulations performed by~\cite{Bonnerot:2020pyz} also show that the circularization of the debris happens on a timescale similar to $t_{\rm fb} \sim 10^7\,{\rm s}$. 
A similar delay timescale for neutrino arrival could be due to a contracting debris configuration~\citep{Metzger:2022lob}.

With three coincident neutrino detections by IceCube, it is timely to look at multi-messenger signatures from TDEs to ascertain and understand the various emission regions and also the complicated dynamics of the debris possibly in the presence of a jet. We presented the upper limit on the total energy of choked jets as a function of $t_{\rm lag}$ in Figure~\ref{fig:nuulplot}, along with the IceCube point-source limit. We found the neutrino association of AT2019dsg could be explained using our delayed choked jet model for $t_{\rm lag} \gtrsim 8 \times 10^6\,{\rm s}$, assuming $v_{\rm deb} = 0.03 c$. Explaining the neutrino observation associated with AT2019fdr is challenging since the required time delay is too long. These conclusions can be further tested with new neutrino associations with TDEs in the upcoming high-energy neutrino observatories like IceCube-Gen2~\citep{IceCube:2014gqr}, KM3NeT~\citep{KM3Net:2016zxf} and electromagnetic telescopes, which would be well-placed to conduct extensive multi-messenger searches that will allow us to better understand the physical mechanism of TDEs.

 \section*{Acknowledgements}
We thank Jose Carpio for useful conversations during the initial stages of this work. We thank Kunihito Ioka and Shigeo S. Kimura for useful comments. We also thank the Institute for Advanced Study (IAS), Princeton for their hospitality where parts of this work was done. M.M and K.M. acknowledge support from  NSF Grant No.~AST-2108466.  
M.B. acknowledges support from the Eberly Postdoctoral Research Fellowship at the Pennsylvania State University. 
The work of K.M. is also supported by the NSF Grant No.~AST-1908689 and No.~AST-2108467, and KAKENHI No.~20H01901 and No.~20H05852.
M.M. also acknowledges support from the Institute for Gravitation and the Cosmos (IGC) Postdoctoral Fellowship. M.M. wishes to thank the Yukawa Institute for Theoretical Physics (YITP), Kyoto University for hospitality where a major part of this work was done.  

\section*{Data availability}
The data underlying this article will be shared on reasonable request to the corresponding author.

\bibliography{refs}

\begin{thebibliography}{}
\makeatletter
\relax
\def\mn@urlcharsother{\let\do\@makeother \do\$\do\&\do\#\do\^\do\_\do\%\do\~}
\def\mn@doi{\begingroup\mn@urlcharsother \@ifnextchar [ {\mn@doi@} {\mn@doi@[]}}
\def\mn@doi@[#1]#2{\def\@tempa{#1}\ifx\@tempa\@empty \href {http://dx.doi.org/#2} {doi:#2}\else \href {http://dx.doi.org/#2} {#1}\fi \endgroup}
\def\mn@eprint#1#2{\mn@eprint@#1:#2::\@nil}
\def\mn@eprint@arXiv#1{\href {http://arxiv.org/abs/#1} {{\tt arXiv:#1}}}
\def\mn@eprint@dblp#1{\href {http://dblp.uni-trier.de/rec/bibtex/#1.xml} {dblp:#1}}
\def\mn@eprint@#1:#2:#3:#4\@nil{\def\@tempa {#1}\def\@tempb {#2}\def\@tempc {#3}\ifx \@tempc \@empty \let \@tempc \@tempb \let \@tempb \@tempa \fi \ifx \@tempb \@empty \def\@tempb {arXiv}\fi \@ifundefined {mn@eprint@\@tempb}{\@tempb:\@tempc}{\expandafter \expandafter \csname mn@eprint@\@tempb\endcsname \expandafter{\@tempc}}}

\bibitem[\protect\citeauthoryear{Aartsen et~al.}{Aartsen et~al.}{2020}]{IceCube:2019cia}
Aartsen M.~G.,  et~al., 2020, \mn@doi [Phys. Rev. Lett.] {10.1103/PhysRevLett.124.051103}, 124, 051103

\bibitem[\protect\citeauthoryear{Adrian-Martinez et~al.}{Adrian-Martinez et~al.}{2016}]{KM3Net:2016zxf}
Adrian-Martinez S.,  et~al., 2016, \mn@doi [J. Phys. G] {10.1088/0954-3899/43/8/084001}, 43, 084001

\bibitem[\protect\citeauthoryear{Alexander, van Velzen, Horesh  \& Zauderer}{Alexander et~al.}{2020}]{Alexander:2020xwb}
Alexander K.~D.,  van Velzen S.,  Horesh A.,   Zauderer B.~A.,  2020, \mn@doi [Space Sci. Rev.] {10.1007/s11214-020-00702-w}, 216, 81

\bibitem[\protect\citeauthoryear{Aloy, Mueller, Ibanez, Marti  \& MacFadyen}{Aloy et~al.}{2000}]{Aloy:1999ai}
Aloy M.~A.,  Mueller E.,  Ibanez J.~M.,  Marti J.~M.,   MacFadyen A.,  2000, \mn@doi [Astrophys. J. Lett.] {10.1086/312537}, 531, L119

\bibitem[\protect\citeauthoryear{Alves~Batista \& Silk}{Alves~Batista \& Silk}{2017}]{AlvesBatista:2017shr}
Alves~Batista R.,  Silk J.,  2017, \mn@doi [Phys. Rev. D] {10.1103/PhysRevD.96.103003}, 96, 103003

\bibitem[\protect\citeauthoryear{Andreoni et~al.}{Andreoni et~al.}{2022}]{Andreoni:2022afu}
Andreoni I.,  et~al., 2022, \mn@doi [Nature] {10.1038/s41586-022-05465-8}, 612, 430

\bibitem[\protect\citeauthoryear{{Begelman} \& {Cioffi}}{{Begelman} \& {Cioffi}}{1989}]{Begelman:1989jp}
{Begelman} M.~C.,  {Cioffi} D.~F.,  1989, \mn@doi [\apjl] {10.1086/185542}, \href {https://ui.adsabs.harvard.edu/abs/1989ApJ...345L..21B} {345, L21}

\bibitem[\protect\citeauthoryear{Bhattacharya, Horiuchi  \& Murase}{Bhattacharya et~al.}{2022}]{Bhattacharya:2021cjc}
Bhattacharya M.,  Horiuchi S.,   Murase K.,  2022, \mn@doi [Mon. Not. Roy. Astron. Soc.] {10.1093/mnras/stac1721}, 514, 6011

\bibitem[\protect\citeauthoryear{{Bhattacharya}, {Carpio}, {Murase}  \& {Horiuchi}}{{Bhattacharya} et~al.}{2023}]{MB_2023}
{Bhattacharya} M.,  {Carpio} J.~A.,  {Murase} K.,   {Horiuchi} S.,  2023, \mn@doi [\mnras] {10.1093/mnras/stad494}, \href {https://ui.adsabs.harvard.edu/abs/2023MNRAS.tmp..504B} {}

\bibitem[\protect\citeauthoryear{Biehl, Boncioli, Lunardini  \& Winter}{Biehl et~al.}{2018}]{Biehl:2017hnb}
Biehl D.,  Boncioli D.,  Lunardini C.,   Winter W.,  2018, \mn@doi [Sci. Rep.] {10.1038/s41598-018-29022-4}, 8, 10828

\bibitem[\protect\citeauthoryear{Blandford \& Rees}{Blandford \& Rees}{1974}]{Blandford:1974zz}
Blandford R.~D.,  Rees M.~J.,  1974, \mn@doi [Mon. Not. Roy. Astron. Soc.] {10.1093/mnras/169.3.395}, 169, 395

\bibitem[\protect\citeauthoryear{Bloom et~al.}{Bloom et~al.}{2011}]{Bloom:2011xk}
Bloom J.~S.,  et~al., 2011, \mn@doi [Science] {10.1126/science.1207150}, 333, 203

\bibitem[\protect\citeauthoryear{Bonnerot \& Lu}{Bonnerot \& Lu}{2020}]{Bonnerot:2019yjb}
Bonnerot C.,  Lu W.,  2020, \mn@doi [Mon. Not. Roy. Astron. Soc.] {10.1093/mnras/staa1246}, 495, 1374

\bibitem[\protect\citeauthoryear{Bonnerot, Price, Lodato  \& Rossi}{Bonnerot et~al.}{2017}]{Bonnerot:2016krr}
Bonnerot C.,  Price D.~J.,  Lodato G.,   Rossi E.~M.,  2017, \mn@doi [Mon. Not. Roy. Astron. Soc.] {10.1093/mnras/stx1210}, 469, 4879

\bibitem[\protect\citeauthoryear{{Bonnerot}, {Lu}  \& {Hopkins}}{{Bonnerot} et~al.}{2021}]{Bonnerot:2020pyz}
{Bonnerot} C.,  {Lu} W.,   {Hopkins} P.~F.,  2021, \mn@doi [\mnras] {10.1093/mnras/stab398}, \href {https://ui.adsabs.harvard.edu/abs/2021MNRAS.504.4885B} {504, 4885}

\bibitem[\protect\citeauthoryear{Bromberg, Nakar, Piran  \& Sari}{Bromberg et~al.}{2011}]{Bromberg:2011fg}
Bromberg O.,  Nakar E.,  Piran T.,   Sari R.,  2011, \mn@doi [Astrophys. J.] {10.1088/0004-637X/740/2/100}, 740, 100

\bibitem[\protect\citeauthoryear{Brown, Levan, Stanway, Tanvir, Cenko, Berger, Chornock  \& Cucchiaria}{Brown et~al.}{2015}]{Brown:2015amy}
Brown G.~C.,  Levan A.~J.,  Stanway E.~R.,  Tanvir N.~R.,  Cenko S.~B.,  Berger E.,  Chornock R.,   Cucchiaria A.,  2015, \mn@doi [Mon. Not. Roy. Astron. Soc.] {10.1093/mnras/stv1520}, 452, 4297

\bibitem[\protect\citeauthoryear{Burrows et~al.}{Burrows et~al.}{2011}]{Burrows:2011dn}
Burrows D.~N.,  et~al., 2011, \mn@doi [Nature] {10.1038/nature10374}, 476, 421

\bibitem[\protect\citeauthoryear{Capanema, Esmaili  \& Murase}{Capanema et~al.}{2020}]{Capanema:2020rjj}
Capanema A.,  Esmaili A.,   Murase K.,  2020, \mn@doi [Phys. Rev. D] {10.1103/PhysRevD.101.103012}, 101, 103012

\bibitem[\protect\citeauthoryear{Cendes, Alexander, Berger, Eftekhari, Williams  \& Chornock}{Cendes et~al.}{2021}]{Cendes:2021bvp}
Cendes Y.,  Alexander K.~D.,  Berger E.,  Eftekhari T.,  Williams P. K.~G.,   Chornock R.,  2021, \mn@doi [Astrophys. J.] {10.3847/1538-4357/ac110a}, 919, 127

\bibitem[\protect\citeauthoryear{Cendes et~al.}{Cendes et~al.}{2022}]{Cendes:2022dia}
Cendes Y.,  et~al., 2022, \mn@doi [Astrophys. J.] {10.3847/1538-4357/ac88d0}, 938, 28

\bibitem[\protect\citeauthoryear{Cenko et~al.}{Cenko et~al.}{2012}]{Cenko:2011ys}
Cenko S.~B.,  et~al., 2012, \mn@doi [Astrophys. J.] {10.1088/0004-637X/753/1/77}, 753, 77

\bibitem[\protect\citeauthoryear{Chornock et~al.}{Chornock et~al.}{2014}]{Chornock:2013jta}
Chornock R.,  et~al., 2014, \mn@doi [Astrophys. J.] {10.1088/0004-637X/780/1/44}, 780, 44

\bibitem[\protect\citeauthoryear{Dai \& Fang}{Dai \& Fang}{2017}]{Dai:2016gtz}
Dai L.,  Fang K.,  2017, \mn@doi [Mon. Not. Roy. Astron. Soc.] {10.1093/mnras/stx863}, 469, 1354

\bibitem[\protect\citeauthoryear{De~Colle \& Lu}{De~Colle \& Lu}{2020}]{DeColle:2019wzp}
De~Colle F.,  Lu W.,  2020, \mn@doi [New Astron. Rev.] {10.1016/j.newar.2020.101538}, 89, 101538

\bibitem[\protect\citeauthoryear{De~Colle, Guillochon, Naiman  \& Ramirez-Ruiz}{De~Colle et~al.}{2012}]{DeColle:2012np}
De~Colle F.,  Guillochon J.,  Naiman J.,   Ramirez-Ruiz E.,  2012, \mn@doi [Astrophys. J.] {10.1088/0004-637X/760/2/103}, 760, 103

\bibitem[\protect\citeauthoryear{Evans \& Kochanek}{Evans \& Kochanek}{1989}]{Evans:1989qe}
Evans C.~R.,  Kochanek C.~S.,  1989, \mn@doi [Astrophys. J. Lett.] {10.1086/168002}, 346, L13

\bibitem[\protect\citeauthoryear{Farrar \& Gruzinov}{Farrar \& Gruzinov}{2009}]{Farrar:2008ex}
Farrar G.~R.,  Gruzinov A.,  2009, \mn@doi [Astrophys. J.] {10.1088/0004-637X/693/1/329}, 693, 329

\bibitem[\protect\citeauthoryear{{Farrar} \& {Piran}}{{Farrar} \& {Piran}}{2014}]{Farrar:2014yla}
{Farrar} G.~R.,  {Piran} T.,  2014, \mn@doi [arXiv e-prints] {10.48550/arXiv.1411.0704}, \href {https://ui.adsabs.harvard.edu/abs/2014arXiv1411.0704F} {p. arXiv:1411.0704}

\bibitem[\protect\citeauthoryear{Generozov, Mimica, Metzger, Stone, Giannios  \& Aloy}{Generozov et~al.}{2017}]{Generozov:2016oon}
Generozov A.,  Mimica P.,  Metzger B.~D.,  Stone N.~C.,  Giannios D.,   Aloy M.~A.,  2017, \mn@doi [Mon. Not. Roy. Astron. Soc.] {10.1093/mnras/stw2439}, 464, 2481

\bibitem[\protect\citeauthoryear{Gezari et~al.}{Gezari et~al.}{2008}]{Gezari:2007bw}
Gezari S.,  et~al., 2008, \mn@doi [Astrophys. J.] {10.1086/529008}, 676, 944

\bibitem[\protect\citeauthoryear{Giannios \& Metzger}{Giannios \& Metzger}{2011}]{Giannios:2011it}
Giannios D.,  Metzger B.~D.,  2011, \mn@doi [Mon. Not. Roy. Astron. Soc.] {10.1111/j.1365-2966.2011.19188.x}, 416, 2102

\bibitem[\protect\citeauthoryear{Golightly, Coughlin  \& Nixon}{Golightly et~al.}{2019}]{Golightly:2019jib}
Golightly E.,  Coughlin E.,   Nixon C.,  2019, \mn@doi [Astrophys. J.] {10.3847/1538-4357/aafd2f}, 872, 163

\bibitem[\protect\citeauthoryear{Gu\'epin, Kotera, Barausse, Fang  \& Murase}{Gu\'epin et~al.}{2018}]{Guepin:2017abw}
Gu\'epin C.,  Kotera K.,  Barausse E.,  Fang K.,   Murase K.,  2018, \mn@doi [Astron. Astrophys.] {10.1051/0004-6361/201732392}, 616, A179

\bibitem[\protect\citeauthoryear{Hamidani \& Ioka}{Hamidani \& Ioka}{2020}]{Hamidani:2020krf}
Hamidani H.,  Ioka K.,  2020, \mn@doi [Mon. Not. Roy. Astron. Soc.] {10.1093/mnras/staa3276}, 500, 627

\bibitem[\protect\citeauthoryear{Hamidani, Kiuchi  \& Ioka}{Hamidani et~al.}{2020}]{Hamidani:2019qyx}
Hamidani H.,  Kiuchi K.,   Ioka K.,  2020, \mn@doi [Mon. Not. Roy. Astron. Soc.] {10.1093/mnras/stz3231}, 491, 3192

\bibitem[\protect\citeauthoryear{Harrison, Gottlieb  \& Nakar}{Harrison et~al.}{2018}]{Harrison:2017jvs}
Harrison R.,  Gottlieb O.,   Nakar E.,  2018, \mn@doi [Mon. Not. Roy. Astron. Soc.] {10.1093/mnras/sty760}, 477, 2128

\bibitem[\protect\citeauthoryear{Hayasaki \& Jonker}{Hayasaki \& Jonker}{2021}]{Hayasaki:2021ggg}
Hayasaki K.,  Jonker P.~G.,  2021, \mn@doi [Astrophys. J.] {10.3847/1538-4357/ac18c2}, 921, 20

\bibitem[\protect\citeauthoryear{{Hayasaki} \& {Yamazaki}}{{Hayasaki} \& {Yamazaki}}{2019}]{Hayasaki:2019kjy}
{Hayasaki} K.,  {Yamazaki} R.,  2019, \mn@doi [\apj] {10.3847/1538-4357/ab44ca}, \href {https://ui.adsabs.harvard.edu/abs/2019ApJ...886..114H} {886, 114}

\bibitem[\protect\citeauthoryear{Hayasaki, Stone  \& Loeb}{Hayasaki et~al.}{2013}]{Hayasaki:2012ia}
Hayasaki K.,  Stone N.,   Loeb A.,  2013, \mn@doi [Mon. Not. Roy. Astron. Soc.] {10.1093/mnras/stt871}, 434, 909

\bibitem[\protect\citeauthoryear{Hayasaki, Stone  \& Loeb}{Hayasaki et~al.}{2016}]{Hayasaki:2015pxa}
Hayasaki K.,  Stone N.~C.,   Loeb A.,  2016, \mn@doi [Mon. Not. Roy. Astron. Soc.] {10.1093/mnras/stw1387}, 461, 3760

\bibitem[\protect\citeauthoryear{Horesh, Cenko  \& Arcavi}{Horesh et~al.}{2021}]{Horesh:2021gvp}
Horesh A.,  Cenko S.~B.,   Arcavi I.,  2021, \mn@doi [Nature Astron.] {10.1038/s41550-021-01300-8}, 5, 491

\bibitem[\protect\citeauthoryear{{IceCube-Gen2 Collaboration} et~al.,}{{IceCube-Gen2 Collaboration} et~al.}{2014}]{IceCube:2014gqr}
{IceCube-Gen2 Collaboration} et~al., 2014, \mn@doi [arXiv e-prints] {10.48550/arXiv.1412.5106}, \href {https://ui.adsabs.harvard.edu/abs/2014arXiv1412.5106I} {p. arXiv:1412.5106}

\bibitem[\protect\citeauthoryear{{Kelley}, {Tchekhovskoy}  \& {Narayan}}{{Kelley} et~al.}{2014}]{Kelly2014}
{Kelley} L.~Z.,  {Tchekhovskoy} A.,   {Narayan} R.,  2014, \mn@doi [\mnras] {10.1093/mnras/stu2041}, \href {https://ui.adsabs.harvard.edu/abs/2014MNRAS.445.3919K} {445, 3919}

\bibitem[\protect\citeauthoryear{Kimura, Murase, Bartos, Ioka, Heng  \& M\'esz\'aros}{Kimura et~al.}{2018}]{Kimura:2018vvz}
Kimura S.~S.,  Murase K.,  Bartos I.,  Ioka K.,  Heng I.~S.,   M\'esz\'aros P.,  2018, \mn@doi [Phys. Rev. D] {10.1103/PhysRevD.98.043020}, 98, 043020

\bibitem[\protect\citeauthoryear{Kobayashi, Meszaros  \& Zhang}{Kobayashi et~al.}{2004}]{Kobayashi:2003yh}
Kobayashi S.,  Meszaros P.,   Zhang B.,  2004, \mn@doi [Astrophys. J. Lett.] {10.1086/381733}, 601, L13

\bibitem[\protect\citeauthoryear{Komossa}{Komossa}{2015}]{Komossa:2015qya}
Komossa S.,  2015, \mn@doi [JHEAp] {10.1016/j.jheap.2015.04.006}, 7, 148

\bibitem[\protect\citeauthoryear{Lazzati \& Begelman}{Lazzati \& Begelman}{2005}]{Lazzati:2005xv}
Lazzati D.,  Begelman M.,  2005, \mn@doi [Astrophys. J.] {10.1086/430877}, 629, 903

\bibitem[\protect\citeauthoryear{Lazzati, Morsony  \& Begelman}{Lazzati et~al.}{2009}]{Lazzati_2009}
Lazzati D.,  Morsony B.~J.,   Begelman M.~C.,  2009, \mn@doi [The Astrophysical Journal] {10.1088/0004-637X/700/1/L47}, 700, L47

\bibitem[\protect\citeauthoryear{Liska, Tchekhovskoy  \& Quataert}{Liska et~al.}{2020}]{Liska:2018btr}
Liska M. T.~P.,  Tchekhovskoy A.,   Quataert E.,  2020, \mn@doi [Mon. Not. Roy. Astron. Soc.] {10.1093/mnras/staa955}, 494, 3656

\bibitem[\protect\citeauthoryear{Liu, Xi  \& Wang}{Liu et~al.}{2020}]{Liu:2020isi}
Liu R.-Y.,  Xi S.-Q.,   Wang X.-Y.,  2020, \mn@doi [Phys. Rev. D] {10.1103/PhysRevD.102.083028}, 102, 083028

\bibitem[\protect\citeauthoryear{Loeb \& Ulmer}{Loeb \& Ulmer}{1997}]{Loeb:1997dv}
Loeb A.,  Ulmer A.,  1997, \mn@doi [Astrophys. J.] {10.1086/304814}, 489, 573

\bibitem[\protect\citeauthoryear{Lu \& Bonnerot}{Lu \& Bonnerot}{2020}]{Lu:2019hwv}
Lu W.,  Bonnerot C.,  2020, \mn@doi [Mon. Not. Roy. Astron. Soc.] {10.1093/mnras/stz3405}, 492, 686

\bibitem[\protect\citeauthoryear{Lunardini \& Winter}{Lunardini \& Winter}{2017}]{Lunardini:2016xwi}
Lunardini C.,  Winter W.,  2017, \mn@doi [Phys. Rev. D] {10.1103/PhysRevD.95.123001}, 95, 123001

\bibitem[\protect\citeauthoryear{Mack, Prieto, Brunetti  \& Orienti}{Mack et~al.}{2009}]{Mack:2008wf}
Mack K.~H.,  Prieto M.~A.,  Brunetti G.,   Orienti M.,  2009, \mn@doi [Mon. Not. Roy. Astron. Soc.] {10.1111/j.1365-2966.2008.14081.x}, 392, 705

\bibitem[\protect\citeauthoryear{Matsumoto \& Piran}{Matsumoto \& Piran}{2021}]{Matsumoto:2021gcg}
Matsumoto T.,  Piran T.,  2021, \mn@doi [Mon. Not. Roy. Astron. Soc.] {10.1093/mnras/stab2418}, 507, 4196

\bibitem[\protect\citeauthoryear{Matsumoto, Piran  \& Krolik}{Matsumoto et~al.}{2022}]{Matsumoto:2021qqo}
Matsumoto T.,  Piran T.,   Krolik J.~H.,  2022, \mn@doi [Mon. Not. Roy. Astron. Soc.] {10.1093/mnras/stac382}, 511, 5085

\bibitem[\protect\citeauthoryear{Matzner}{Matzner}{2003}]{Matzner:2002ti}
Matzner C.~D.,  2003, \mn@doi [Mon. Not. Roy. Astron. Soc.] {10.1046/j.1365-8711.2003.06969.x}, 345, 575

\bibitem[\protect\citeauthoryear{Matzner \& McKee}{Matzner \& McKee}{1999}]{Matzner:1998mg}
Matzner C.~D.,  McKee C.~F.,  1999, \mn@doi [Astrophys. J.] {10.1086/306571}, 510, 379

\bibitem[\protect\citeauthoryear{Meszaros \& Waxman}{Meszaros \& Waxman}{2001}]{Meszaros:2001ms}
Meszaros P.,  Waxman E.,  2001, \mn@doi [Phys. Rev. Lett.] {10.1103/PhysRevLett.87.171102}, 87, 171102

\bibitem[\protect\citeauthoryear{Metzger}{Metzger}{2022}]{Metzger:2022lob}
Metzger B.~D.,  2022, \mn@doi [Astrophys. J. Lett.] {10.3847/2041-8213/ac90ba}, 937, L12

\bibitem[\protect\citeauthoryear{Mimica, Giannios, Metzger  \& Aloy}{Mimica et~al.}{2015}]{Mimica:2015qka}
Mimica P.,  Giannios D.,  Metzger B.~D.,   Aloy M.~A.,  2015, \mn@doi [Mon. Not. Roy. Astron. Soc.] {10.1093/mnras/stv825}, 450, 2824

\bibitem[\protect\citeauthoryear{Mizuta \& Aloy}{Mizuta \& Aloy}{2009}]{Mizuta:2008ch}
Mizuta A.,  Aloy M.~A.,  2009, \mn@doi [Astrophys. J.] {10.1088/0004-637X/699/2/1261}, 699, 1261

\bibitem[\protect\citeauthoryear{Mizuta \& Ioka}{Mizuta \& Ioka}{2013}]{Mizuta:2013yma}
Mizuta A.,  Ioka K.,  2013, \mn@doi [Astrophys. J.] {10.1088/0004-637X/777/2/162}, 777, 162

\bibitem[\protect\citeauthoryear{Murase}{Murase}{2008}]{Murase:2008zzc}
Murase K.,  2008, \mn@doi [AIP Conf. Proc.] {10.1063/1.3027912}, 1065, 201

\bibitem[\protect\citeauthoryear{Murase, Guetta  \& Ahlers}{Murase et~al.}{2016}]{Murase:2015xka}
Murase K.,  Guetta D.,   Ahlers M.,  2016, \mn@doi [Phys. Rev. Lett.] {10.1103/PhysRevLett.116.071101}, 116, 071101

\bibitem[\protect\citeauthoryear{Murase, Kimura, Zhang, Oikonomou  \& Petropoulou}{Murase et~al.}{2020}]{Murase:2020lnu}
Murase K.,  Kimura S.~S.,  Zhang B.~T.,  Oikonomou F.,   Petropoulou M.,  2020, \mn@doi [Astrophys. J.] {10.3847/1538-4357/abb3c0}, 902, 108

\bibitem[\protect\citeauthoryear{Nagakura, Ito, Kiuchi  \& Yamada}{Nagakura et~al.}{2011}]{Nagakura:2010zt}
Nagakura H.,  Ito H.,  Kiuchi K.,   Yamada S.,  2011, \mn@doi [Astrophys. J.] {10.1088/0004-637X/731/2/80}, 731, 80

\bibitem[\protect\citeauthoryear{{Nakar} \& {Granot}}{{Nakar} \& {Granot}}{2007}]{NakarGranot2007}
{Nakar} E.,  {Granot} J.,  2007, \mn@doi [\mnras] {10.1111/j.1365-2966.2007.12245.x}, \href {https://ui.adsabs.harvard.edu/abs/2007MNRAS.380.1744N} {380, 1744}

\bibitem[\protect\citeauthoryear{Narayan, Chael, Chatterjee, Ricarte  \& Curd}{Narayan et~al.}{2022}]{Narayan:2021qfw}
Narayan R.,  Chael A.,  Chatterjee K.,  Ricarte A.,   Curd B.,  2022, \mn@doi [Mon. Not. Roy. Astron. Soc.] {10.1093/mnras/stac285}, 511, 3795

\bibitem[\protect\citeauthoryear{Panaitescu \& Kumar}{Panaitescu \& Kumar}{2000}]{Panaitescu:2000bk}
Panaitescu A.,  Kumar P.,  2000, \mn@doi [Astrophys. J.] {10.1086/317090}, 543, 66

\bibitem[\protect\citeauthoryear{Park, Caprioli  \& Spitkovsky}{Park et~al.}{2015}]{Park:2014lqa}
Park J.,  Caprioli D.,   Spitkovsky A.,  2015, \mn@doi [Phys. Rev. Lett.] {10.1103/PhysRevLett.114.085003}, 114, 085003

\bibitem[\protect\citeauthoryear{Phinney}{Phinney}{1989}]{Phinney1989}
Phinney E.~S.,  1989, in Morris M.,  ed., The Center of the Galaxy. Springer Netherlands, Dordrecht, pp 543--553

\bibitem[\protect\citeauthoryear{Piran, Svirski, Krolik, Cheng  \& Shiokawa}{Piran et~al.}{2015}]{Piran:2015gha}
Piran T.,  Svirski G.,  Krolik J.,  Cheng R.~M.,   Shiokawa H.,  2015, \mn@doi [Astrophys. J.] {10.1088/0004-637X/806/2/164}, 806, 164

\bibitem[\protect\citeauthoryear{Rees}{Rees}{1988}]{Rees:1988bf}
Rees M.~J.,  1988, \mn@doi [Nature] {10.1038/333523a0}, 333, 523

\bibitem[\protect\citeauthoryear{Reusch et~al.}{Reusch et~al.}{2022}]{Reusch:2021ztx}
Reusch S.,  et~al., 2022, \mn@doi [Phys. Rev. Lett.] {10.1103/PhysRevLett.128.221101}, 128, 221101

\bibitem[\protect\citeauthoryear{Ryu, Krolik, Piran  \& Noble}{Ryu et~al.}{2020}]{Ryu:2020huz}
Ryu T.,  Krolik J.,  Piran T.,   Noble S.~C.,  2020, \mn@doi [Astrophys. J.] {10.3847/1538-4357/abb3cd}, 904, 99

\bibitem[\protect\citeauthoryear{Senno, Murase  \& Meszaros}{Senno et~al.}{2017}]{Senno:2016bso}
Senno N.,  Murase K.,   Meszaros P.,  2017, \mn@doi [Astrophys. J.] {10.3847/1538-4357/aa6344}, 838, 3

\bibitem[\protect\citeauthoryear{Stein}{Stein}{2020}]{Stein:2019ivm}
Stein R.,  2020, \mn@doi [PoS] {10.22323/1.358.1016}, ICRC2019, 1016

\bibitem[\protect\citeauthoryear{Stein et~al.}{Stein et~al.}{2021}]{Stein:2020xhk}
Stein R.,  et~al., 2021, \mn@doi [Nature Astron.] {10.1038/s41550-020-01295-8}, 5, 510

\bibitem[\protect\citeauthoryear{Stern et~al.}{Stern et~al.}{2004}]{Stern:2004xb}
Stern D.,  et~al., 2004, \mn@doi [Astrophys. J.] {10.1086/422744}, 612, 690

\bibitem[\protect\citeauthoryear{Stone, Sari  \& Loeb}{Stone et~al.}{2013}]{Stone:2012uk}
Stone N.,  Sari R.,   Loeb A.,  2013, \mn@doi [Mon. Not. Roy. Astron. Soc.] {10.1093/mnras/stt1270}, 435, 1809

\bibitem[\protect\citeauthoryear{Strubbe \& Quataert}{Strubbe \& Quataert}{2009}]{Strubbe:2009qs}
Strubbe L.~E.,  Quataert E.,  2009, \mn@doi [Mon. Not. Roy. Astron. Soc.] {10.1111/j.1365-2966.2009.15599.x}, 400, 2070

\bibitem[\protect\citeauthoryear{{Tchekhovskoy}, {Metzger}, {Giannios}  \& {Kelley}}{{Tchekhovskoy} et~al.}{2014}]{Tchekovsky2014}
{Tchekhovskoy} A.,  {Metzger} B.~D.,  {Giannios} D.,   {Kelley} L.~Z.,  2014, \mn@doi [\mnras] {10.1093/mnras/stt2085}, \href {https://ui.adsabs.harvard.edu/abs/2014MNRAS.437.2744T} {437, 2744}

\bibitem[\protect\citeauthoryear{Tingay, Lenc, Brunetti  \& Bondi}{Tingay et~al.}{2008}]{Tingay:2008sg}
Tingay S.~J.,  Lenc E.,  Brunetti G.,   Bondi M.,  2008, \mn@doi [Astron. J.] {10.1088/0004-6256/136/6/2473}, 136, 2473

\bibitem[\protect\citeauthoryear{Van~Velzen et~al.}{Van~Velzen et~al.}{2019}]{vanVelzen:2018dwv}
Van~Velzen S.,  et~al., 2019, \mn@doi [Astrophys. J.] {10.3847/1538-4357/aafe0c}, 872, 198

\bibitem[\protect\citeauthoryear{{Van Velzen} et~al.,}{{Van Velzen} et~al.}{2021}]{vanVelzen:2021zsm}
{Van Velzen} S.,  et~al., 2021, \mn@doi [arXiv e-prints] {10.48550/arXiv.2111.09391}, \href {https://ui.adsabs.harvard.edu/abs/2021arXiv211109391V} {p. arXiv:2111.09391}

\bibitem[\protect\citeauthoryear{Wang \& Liu}{Wang \& Liu}{2016}]{Wang:2015mmh}
Wang X.-Y.,  Liu R.-Y.,  2016, \mn@doi [Phys. Rev. D] {10.1103/PhysRevD.93.083005}, 93, 083005

\bibitem[\protect\citeauthoryear{Wang, Liu, Dai  \& Cheng}{Wang et~al.}{2011}]{Wang:2011ip}
Wang X.-Y.,  Liu R.-Y.,  Dai Z.-G.,   Cheng K.~S.,  2011, \mn@doi [Phys. Rev. D] {10.1103/PhysRevD.84.081301}, 84, 081301

\bibitem[\protect\citeauthoryear{Winter \& Lunardini}{Winter \& Lunardini}{2021}]{Winter:2020ptf}
Winter W.,  Lunardini C.,  2021, \mn@doi [Nature Astron.] {10.1038/s41550-021-01343-x}, 5, 472

\bibitem[\protect\citeauthoryear{Winter \& Lunardini}{Winter \& Lunardini}{2023}]{Winter:2022fpf}
Winter W.,  Lunardini C.,  2023, \mn@doi [Astrophys. J.] {10.3847/1538-4357/acbe9e}, 948, 42

\bibitem[\protect\citeauthoryear{{Yuan} \& {Winter}}{{Yuan} \& {Winter}}{2023}]{Yuan:2023cmd}
{Yuan} C.,  {Winter} W.,  2023, \mn@doi [arXiv e-prints] {10.48550/arXiv.2306.15659}, \href {https://ui.adsabs.harvard.edu/abs/2023arXiv230615659Y} {p. arXiv:2306.15659}

\bibitem[\protect\citeauthoryear{Zauderer et~al.}{Zauderer et~al.}{2011}]{Zauderer:2011mf}
Zauderer B.~A.,  et~al., 2011, \mn@doi [Nature] {10.1038/nature10366}, 476, 425

\bibitem[\protect\citeauthoryear{Zhang, Woosley  \& Heger}{Zhang et~al.}{2004}]{Zhang:2003rp}
Zhang W.-Q.,  Woosley S.~E.,   Heger A.,  2004, \mn@doi [Astrophys. J.] {10.1086/386300}, 608, 365

\bibitem[\protect\citeauthoryear{Zhang, Murase, Oikonomou  \& Li}{Zhang et~al.}{2017}]{Zhang:2017hom}
Zhang B.~T.,  Murase K.,  Oikonomou F.,   Li Z.,  2017, \mn@doi [Phys. Rev. D] {10.1103/PhysRevD.96.063007}, 96, 063007

\bibitem[\protect\citeauthoryear{{Zheng}, {Liu}  \& {Wang}}{{Zheng} et~al.}{2023}]{Zheng:2022kam}
{Zheng} J.-H.,  {Liu} R.-Y.,   {Wang} X.-Y.,  2023, \mn@doi [\apj] {10.3847/1538-4357/ace71c}, \href {https://ui.adsabs.harvard.edu/abs/2023ApJ...954...17Z} {954, 17}

\makeatother
\end{thebibliography}
\label{lastpage}

\end{document}